\documentclass[a4paper,twocolumn,11pt,accepted=2022-09-12]{quantumarticle}
\pdfoutput=1
\usepackage[utf8]{inputenc}
\usepackage[english]{babel}
\usepackage[T1]{fontenc}

\usepackage{graphicx}
\usepackage{dcolumn}
\usepackage{bm}
\usepackage{braket}
\usepackage{placeins}
\usepackage{dsfont}
\usepackage{multirow}
\usepackage{array}
\usepackage[table]{xcolor}
\usepackage{verbatim}
\usepackage{amsmath,amssymb,mathtools}

\usepackage{amsthm}
\theoremstyle{plain}

\usepackage[numbers,square,sort&compress]{natbib}

\definecolor{myurlcolor}{rgb}{0,0,0.7}
\definecolor{myrefcolor}{rgb}{0.8,0,0}
\usepackage[unicode=true,pdfusetitle, bookmarks=false,bookmarksnumbered=false, bookmarksopen=false, breaklinks=false,pdfborder={0 0 0},backref=false, colorlinks=true, linkcolor=myrefcolor,citecolor=myurlcolor,urlcolor=myurlcolor]{hyperref}

\usepackage{tocloft}

\usepackage{etoolbox}
\makeatletter
\newcommand{\mainmatter}{%
	\setcounter{footnote}{0}%
	\patchcmd{\@makefntext}{\fnsymbol}{\arabic}{}{}%
	\patchcmd{\@thefnmark}{\fnsymbol}{\arabic}{}{}%
	\def\@makefnmark{\textsuperscript{\arabic{footnote}}}%
}
\makeatother

\DeclareGraphicsExtensions{.png,.pdf,.eps,.jpg}

\makeatletter

\newcommand{\trace}{\mathrm{Tr}}

\usepackage{xparse}
\NewDocumentCommand\Tr{g}{
  \IfNoValueTF{#1}
    {\trace}
    {\trace\!\left\{#1\right\}}
}

\newcommand{\proj}[1]{\ket{#1}\!\bra{#1}}

\def\dd{\mathrm d}
\def\ii{\mathrm i}

\newcommand{\trm}[1]{\textrm{#1}}
\newcommand{\mrm}[1]{\mathrm{#1}}

\newcommand{\eref}[1]{(\ref{#1})}
\newcommand{\erefs}[2]{(\ref{#1}-\ref{#2})}
\newcommand{\eqnref}[1]{Eq.~(\ref{#1})}
\newcommand{\eqnsref}[2]{Eqs.~(\ref{#1}-\ref{#2})}
\newcommand{\figref}[1]{Fig.~\ref{#1}}

\newcommand{\secref}[1]{Sec.~\ref{#1}}
\newcommand{\appref}[1]{App.~\ref{#1}}

\newcommand{\citeref}[1]{Ref.~\cite{#1}}

\newcommand{\kappad}{{\kappa_{\mrm{d}}}}
\newcommand{\kappal}{{\kappa_{\mrm{l}}}}

\newcommand{\nph}{\text{no-ph}}
\newcommand{\cav}[1]{{#1}_\text{cav}}
\newcommand{\mech}[1]{{#1}_\text{mech}}

\newcommand{\est}[1]{\tilde{#1}}
\newcommand{\estB}[1]{\tilde{#1}_\text{\tiny{opt}}}

\NewDocumentCommand\av{m+g}{%
  \IfNoValueTF{#2}
    {\left\langle{#1}\right\rangle}
    {\left\langle{#1}\right\rangle_{#2}}%
}
\NewDocumentCommand\Var{m+g}{%
  \IfNoValueTF{#2}
    {\mathrm{Var}\!\left[#1\right]}
    {\mathrm{Var}\!\left[#1\right]_{#2}}%
}

\newcommand{\eqdef}{\coloneqq}

\newcommand{\cond}[1]{{#1}_\mathrm{c}}
\newcommand{\rhoc}{\cond{\rho}}
\newcommand{\psic}{\cond{\psi}}
\newcommand{\kB}{k_\text{B}}

\newcommand{\Hrf}{H_\text{rf}} 
\newcommand{\Hpolaron}{\tilde{H}_\text{rf}} 

\newcommand{\omegaM}{\omega_{\rm M}}
\newcommand{\omegaL}{\omega_{\rm L}}

\makeatother

\begin{document}
	
	\title{%
	Exploiting non-linear effects in optomechanical sensors with continuous photon-counting
	}
	
	\author{Lewis A.~Clark}
	\affiliation{Centre for Quantum Optical Technologies, Centre of New Technologies, University of Warsaw, Banacha 2c, 02-097 Warszawa, Poland}
	\author{Bartosz Markowicz}
	\affiliation{Centre for Quantum Optical Technologies, Centre of New Technologies, University of Warsaw, Banacha 2c, 02-097 Warszawa, Poland}
	\affiliation{Faculty of Physics, University of Warsaw, Pasteura 5, 02-093 Warszawa, Poland}
	\author{Jan Ko\l{}ody\'nski}
	\affiliation{Centre for Quantum Optical Technologies, Centre of New Technologies, University of Warsaw, Banacha 2c, 02-097 Warszawa, Poland}

	\begin{abstract}
	Optomechanical systems are rapidly becoming one of the most promising platforms for observing quantum behaviour, especially at the macroscopic level. Moreover, thanks to their state-of-the-art methods of fabrication, they may now enter regimes of non-linear interactions between their constituent mechanical and optical degrees of freedom. In this work, we show how this novel opportunity may serve to construct a new generation of optomechanical sensors. We consider the canonical optomechanical setup with the detection scheme being based on time-resolved counting of photons leaking from the cavity. By performing simulations and resorting to Bayesian inference, we demonstrate that the non-classical correlations of the detected photons may crucially enhance the sensor performance in real time. We believe that our work may stimulate a new direction in the design of such devices, while our methods apply also to other platforms exploiting non-linear light-matter interactions and photon detection.
	\end{abstract}

	\section{Introduction}
	\label{sec:intro}
	Since the quantisation of the interaction between an optical and mechanical mode~\cite{Law1995}, quantum optomechanics~\cite{Aspelmeyer2014,Aspelmeyer2014a,Bowen2015} has led to numerous experimental breakthroughs~\cite{barzanjeh_optomechanics_2022}, summarised spectacularly by the recent achievement of cooling a 10-kg object close to its motional ground state~\cite{whittle_approaching_2021}. Its underlying theoretical framework, despite originating from moving-end optical cavities~\cite{Mancini1997,Bose1997,Clerk2014}, has been successfully shown to apply to a variety of systems, such as:~levitated nanoparticles~\cite{Gonzalez2021,tebbenjohanns_quantum_2021,kiesel_cavity_2013}, trapped ultracold atoms~\cite{Brennecke2008,murch_observation_2008,brooks_non-classical_2012}, photonic crystals~\cite{eichenfield_optomechanical_2009,Chan2011,riedinger_remote_2018} or optical microresonators~\cite{armani_ultra-high-q_2003,wilson_measurement-based_2015,sudhir_appearance_2017}. 

	Furthermore, owing to the rapid advancement of their detection schemes, optomechanical devices were demonstrated to be controllable in real time, not only for ground-state feedback cooling~\cite{rossi_measurement-based_2018,whittle_approaching_2021,wilson_measurement-based_2015}, but also for continuous tracking of their micromechanical motion~\cite{Iwasawa2013,wieczorek_optimal_2015,rossi_observing_2019}, allowing for applicability of real-time feedback to attain the optimal measurement sensitivity~\cite{sudhir_appearance_2017,Setter2018,mason_continuous_2019,magrini_real-time_2021}.

	Still, the vast majority of these experiments were conducted within the so-called linearised regime of interactions~\cite{vitali_optomechanical_2007,Genes2008,WilsonRae2008,Liu2013}, which facilitates the description -- with the framework of Gaussian states/measurements~\cite{ferraro_gaussian_2005} and feedback~\cite{hofer_chapter_2017} being then applicable. In contrast, considering the novel optomechanical setups that exhibit strong single-photon coupling, e.g.~ones involving ultracold atoms~\cite{Brennecke2008,murch_observation_2008} or hybrid devices achieving non-linearity by employing an auxiliary system~\cite{oconnell_quantum_2010,Stannigel2012,ramos_nonlinear_2013,reed_faithful_2017}, but also anticipating their advent within other platforms~\cite{Teufel2011}, one must return to the exact solutions of system dynamics~\cite{Mancini1997,Bose1997} that, however, are analytically tractable only if particular limited forms of decoherence and optical driving are accounted for~\cite{Qvarfort2021} -- unless one resorts to numerical methods~\cite{Wang2011,Bergholm2019}.

	Importantly, it is the non-linear effects that allow a single photon to be converted into multiple phonons and vice versa, so that phenomena such as blockades~\cite{Nunnenkamp2011,Rabl2011} or cascades~\cite{Xu2013} of photons then become possible, with the light leaving the cavity exhibiting a clear non-classical character~\cite{Kronwald2013}. In particular, as the leaking of these photons may then contain more information about the parameters of the optomechanical device due to their correlations, such a system becomes very sensitive to external perturbations~\cite{Clark2019} -- constituting a model \emph{sensor}. On the other hand, the positive impact of non-linearity on the sensitivity of optomechanical devices have been recently observed in the limit of asymptotic statistics~\cite{qvarfort_gravimetry_2018,qvarfort_optimal_2021}, i.e.~in tasks where the same preparation-measurement procedure may be repeated arbitrary amount of times, so that the tools of frequentist estimation theory such as Fisher information~\cite{Kay1993}, and its quantum generalisations~\cite{Paris2009}, become applicable.

	In this work, we focus on sensing tasks in which the optomechanical device is monitored in \emph{real time} -- as motivated by the linear-regime experiments~\cite{wieczorek_optimal_2015,rossi_observing_2019} -- so that the data, which is then continuously gathered in a single experimental run, must be efficiently used to infer the parameter being sensed. Moreover, in order to be able to benefit from the non-linear effects, we consider the photon-counting detection scheme -- see e.g.~\cite{cohen_phonon_2015,galinskiy_phonon_2020,fiaschi_optomechanical_2021} for recent implementations -- for which the information about the parameter is then contained within the (non-classical) photon-click patterns being recorded~\cite{Kronwald2013}. In particular, we resort to continuous measurement theory~\cite{Jacobs2014}, in order to firstly demonstrate that thanks to continuous monitoring the resulting conditional (non-linear) dynamics of the optomechanical sensor exhibits enhanced \emph{entanglement} between optical and mechanical degrees of freedom, despite both photonic and phononic dissipation. We then inspect explicitly the time-correlations of the detected photons, i.e.~the corresponding second-order correlation functions, in order to identify and focus on the three distinct regimes of photon blockade~\cite{Rabl2011}, cascade~\cite{Xu2013} and on-resonance optical driving, which are controlled by adjusting the (red) detuning of the driving field~\cite{Kronwald2013}. 

	Working with low photon and phonon excitation numbers, we are able to efficiently simulate the non-linear open-system dynamics and, in particular, use Bayesian inference theory~\cite{Gammelmark2013} to construct real-time estimators for the parameter of interest. Although in our study we focus on estimating the optomechanical coupling strength~\cite{Bernad2018}, our approach can be applied to any parameter of the system affecting the photon-click statistics. We also show this explicitly for the regime of on-resonance driving in \appref{app:sensing_simple}, where we infer the frequency of the mechanical oscillator instead -- a task motivated by applications in scanning force microscopy~\cite{halg_membrane-based_2021}.

	Nonetheless, in order to assess the metrological capabilities of the sensor's conditional evolution, we firstly compute the ultimate bounds on achievable sensitivity that account for the imperfect \emph{a priori} knowledge about the parameter, the Van Trees bounds~\cite{van2007bayesian}, along particular photon-click trajectories; and compare them with the unconditional (ensemble average) case. Being based on quantum Fisher Information~\cite{Paris2009}, these constitute then a benchmark for the idealistic single-shot measurement that would have to be destructively performed at a given time, while crucially ignoring the information contained within the photon-clicks previously registered~\cite{albarelli_ultimate_2017}. We show that, although the precision of such a measurement is enhanced along a conditional trajectory, the decoherence inevitably drives the sensor towards a stationary state and, hence, puts then a fundamental limit on the attainable sensitivity. 

	As we demonstrate, this is in stark contrast to the experimentally motivated setting in which one has access ``only'' to the photon-click pattern. Thanks to optical driving, the photons continuously leak from the cavity despite dissipation, while the subsequent detections reveal more and more information about the unknown parameter, which can then in principle be inferred up to any desired precision, given a single but sufficiently long run of the experiment. Strikingly, we observe that although less photons are typically emitted in the photon-cascade regime~\cite{Xu2013}, the emitted photons may yield much higher sensitivities due to their strong non-classical correlations. As a consequence, we believe that our results pave the way for novel applications of non-linear effects in optomechanical sensors, stimulating future proposals to investigate the impact of non-linearity on schemes involving, e.g.,~active feedback~\cite{Bergholm2019}. Moreover, our analysis based on Bayesian inference~\cite{kiilerich_estimation_2014} could in future also be conducted for other platforms involving light-matter interfaces, whose non-linearity can be engineered to tune the non-classical properties of photons being emitted~\cite{chang_quantum_2014}, for example with single~\cite{Reiserer2015} or Rydberg~\cite{peyronel_quantum_2012,Mohl2020} atoms coupled to a cavity, or a waveguide~\cite{prasad_correlating_2020}.
	
	The manuscript is organised as follows.  In \secref{sec:dynam}, we review the required theory of quantum optomechanics and provide solutions to the dynamics both in their closed and open form, including the stochastic unravelling of the latter that allows us to model continuous photon-detection. We then study in \secref{sec:entanglement} how the conditional evolution along a given trajectory (determined by a continuous-measurement record) enhances the entanglement between the optical and mechanical degrees of freedom. In \secref{sec:photon_statisics}, we analyse the simulated statistics of photon-click patterns and evaluate the corresponding second-order correlation functions, in order to verify how the system characteristics affect the non-classical correlations of detected photons. We turn to the theory of real-time quantum sensing in \secref{sec:bayesian}, where we present the tools of Bayesian inference, including the ultimate quantum bounds on the average precision. Finally, in \secref{sec:sensing} we demonstrate how the Bayesian formulation allows us to estimate accurately the unknown parameter from photon-click patterns, while benefiting from the non-linear affects. We end with conclusions in \secref{sec:conclusions}.

	\section{Optomechanical system} 
	\label{sec:dynam}
	
	\begin{figure*}[t!] 
		\centering
		\includegraphics[width=0.8\linewidth]{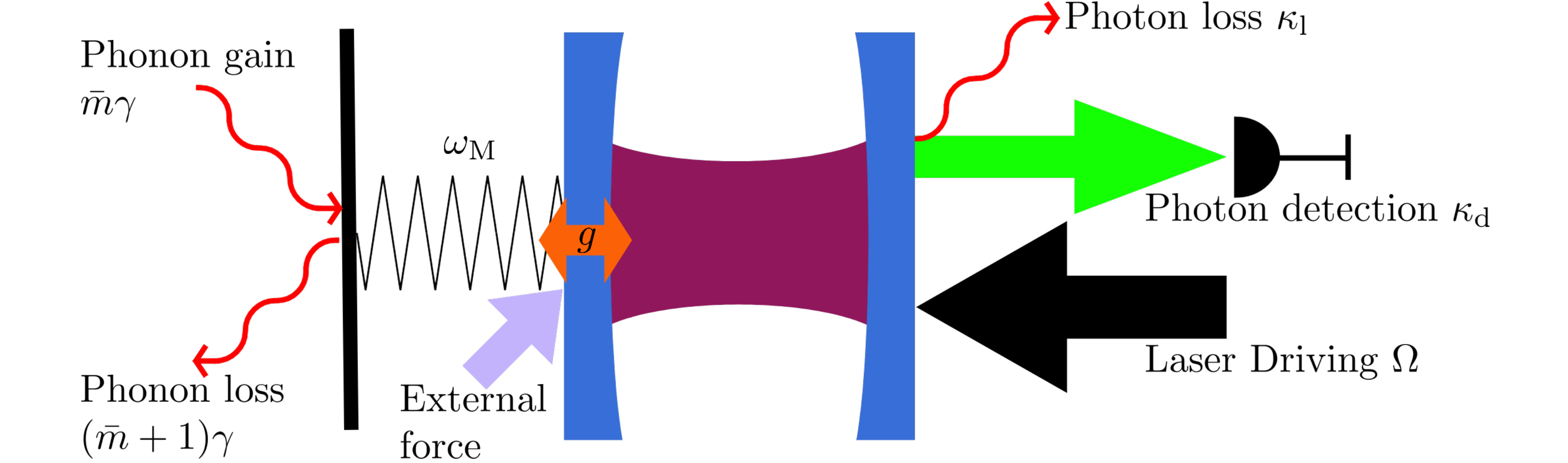}
		\caption{ 
			\textbf{Optomechanical sensor.}
			An optomechanical cavity is driven with laser Rabi frequency $\Omega$. The optical field created inside the cavity couples to the mechanical mode with strength $g$. The photons leak out of the cavity and are detected with the rate $\kappad$, while further photons are lost at a rate $\kappal$. The phonons contained within the mechanical oscillator are both excited and lost due to interactions with a thermal reservoir at rates $\gamma(\bar{m}+1)$ and $\gamma\bar{m}$ respectively, with $\bar{m}$ parametrising effectively the temperature of the bath. External force to be sensed affects the mechanical oscillator varying its light-coupling strength $g$, natural frequency $\omegaM$, or both.
		}
		\label{Fig:Cavity1}
	\end{figure*}
	
	We shall consider the canonical optomechanical system depicted in \figref{Fig:Cavity1}, in the form of a laser-driven optical cavity with one mirror free to oscillate~\cite{Aspelmeyer2014}.  We assume monochromatic driving at frequency $\omegaL$, and interactions with a single mechanical mode of natural frequency $\omegaM$, whose oscillations are small compared to the cavity's unperturbed length.  The Hamiltonian describing such a system with the cavity frequency expanded up to first order is reads~\cite{Mancini1997,Bose1997,Clerk2014}:
	\begin{align}
		\label{Eq:H}
		H = &\; \hbar \omega_0 a^\dagger a + \hbar \omegaM b^\dagger b + \hbar G x a^\dagger a \nonumber \\
		& + \frac{\hbar}{2} \left( \Omega e^{{\rm i} \omegaL t} a +  \Omega^* e^{-{\rm i} \omegaL t} a^\dagger \right) \, ,
	\end{align}
	where $a$ ($a^\dagger$) and $b$ ($b^\dagger$) are the bosonic annihilation (creation) operators of the cavity and mechanical modes, respectively, while $\omega_0$ denotes the unperturbed cavity frequency, and $G$ is the optomechanical frequency shift per displacement~\cite{Clerk2014}. The strength of the laser driving is then parametrised by the Rabi frequency $\Omega$.  Finally, $x$ is the position operator of the mechanics, given by $x = x_{\rm ZPF} (b + b^\dagger)$, where $x_{\rm ZPF} = \sqrt{\hbar/2 m \omegaM}$ is the mechanical zero point fluctuation for an oscillator of mass $m$.  Defining the optomechanical coupling as $g = G x_{\rm ZPF}$, and moving into the rotating frame with respect to the laser frequency, we may rewrite the Hamiltonian \eref{Eq:H} as
	\begin{align} 
		\label{Eq:H_RF}
		\Hrf = & - \hbar \Delta a^\dagger a + \hbar \omegaM b^\dagger b + \hbar g a^\dagger a (b + b^\dagger) \nonumber \\
		& + \frac{\hbar}{2} \left(\Omega a + \Omega^* a^\dagger \right) \, ,
	\end{align}
	with $\Delta = \omegaL - \omega_0$ parametrising then the detuning of the driving laser.  
	
	At this point, the Hamiltonian is often linearised by assuming further the light to leak from the cavity at a sufficiently large overall rate $\kappa$.  Here though, we are interested in the parameter regime where the non-linear effects are important, corresponding to the parameters satisfying~\cite{Aspelmeyer2014}:
	\begin{align} \label{Eq:Non-lin}
		\frac{g}{\kappa} & > 1 \, , \quad \frac{g^2}{\omegaM \kappa} > 1 \, .
	\end{align}	
	Hence, without making any approximations -- in particular, the linearisation -- we transform the Hamiltonian \eref{Eq:H_RF}
	(via the so-called polaron transform) onto $\Hpolaron=U^\dagger \Hrf U$ with the unitary operator $U = \exp\!\left[g a^\dagger a \left(b^\dagger - b \right) / \omegaM \right]$, so that \eqnref{Eq:H_RF} now reads~\cite{Aspelmeyer2014}:
	\begin{align} 
	\label{Eq:H_pol}
		\Hpolaron = & - \hbar \Delta a^\dagger a + \hbar \omegaM b^\dagger b - \hbar \frac{g^2}{\omegaM} \left(a^\dagger a \right)^2 \nonumber \\
		& + \frac{\hbar}{2} \left(\Omega a D + {\rm H.c.} \right) \, ,
	\end{align}
	where by $D = \exp \left(g \left(b^\dagger - b \right) / \omegaM \right)$ we denote a mechanical displacement (polaron) operator.  In this picture, the cavity and mechanical modes become decoupled at a price of introducing a fictitious Kerr-like interaction between the photons.

	Importantly, the Hamiltonian \eref{Eq:H_pol} allows one to directly deduce the energy-level structure for the non-linear optomechanical system as~\cite{Aspelmeyer2014}:
	\begin{align} 
	\label{Eq:energy_levels}
		E(n_\trm{cav},n_\trm{mech}) = & - \hbar \Delta \,n_\trm{cav} + \hbar \omegaM \,n_\trm{mech} \nonumber \\
		& - \hbar \frac{g^2}{\omegaM} n_\trm{cav}^2 \, ,
	\end{align}
	where $n_\trm{mech}$ ($n_\trm{cav}$) denotes the number of phonons (photons) excited within the mechanical oscillator (cavity). It then follows that, thanks to the quadratic term $\propto n_\trm{cav}^2$, by adjusting the detuning $\Delta$, multi- (rather than single-) photon transitions may become energetically favourable. In particular, by setting $\Delta = - n g^2 / \omegaM$ with $n\in\mathbb{Z}^+$, so that the first and third terms in \eqnref{Eq:energy_levels} cancel at $n_\trm{cav}=n$, the transition $\ket{0}_\trm{cav} \rightarrow \ket{n}_\trm{cav}$ becomes preferable with $n$ photons being excited at once via driving. In what follows, we focus on the three important cases of such a detuning scheme with $n=\{0,1,2\}$, each of which corresponds then to the regime of \emph{on-resonance} driving, a photon \emph{blockade}~\cite{Rabl2011,Nunnenkamp2011} or a ($2$-photon) \emph{cascade}~\cite{Xu2013}, respectively.

	As shown in \figref{Fig:Cavity1}, in order to access the non-classical photon statistics generated by the non-linear interaction, we consider a time-resolved photodection of photons leaking from the cavity at an overall rate $\kappa=\kappad+\kappal$, with $\kappad$ and $\kappal$ accounting for the fraction being actually detected or lost, respectively. We assume the full spectrum of emitted light to be detected~\cite{Kronwald2013}, in particular, without distinguishing photons scattered into different frequency-resolved sidebands. Although multimode resolution may be used to enhance non-classical effects~\cite{genes_robust_2008,Schmidt2021}, opening doors for multimodal engineering of also the driving field~\cite{Boerkje2021}, in our work we consider relatively weak driving fields that result in an already small number of emitted photons, which we do not want to further decrease by photon categorisation.  Ultimately, we are interested in performing sensing with the system.  As pictured in Fig.~\ref{Fig:Cavity1}, we can envisage the scenario of an external force perturbing the system, which may affect its parameters such as $g$ or $\omegaM$, the effect of which should be visible within the observed photon statistics.

	\subsection{Closed dynamics}
	In the case of no driving ($\Omega=0$) and absence of any dissipation, the evolution of the optomechanical system can be described by a unitary time-evolution operator decomposed as follows~\cite{Mancini1997,Bose1997}:
	\begin{align}
		\label{Eq:unitary_evolution_operator}
		U(t') = &\; {\rm e}^{-{\rm i} r a^{\dagger} a t'} {\rm e}^{{\rm i} k^2 \left(a^{\dagger} a\right)^2 \left(t' - \sin (t')\right)} \nonumber \\
		& \times {\rm e}^{k a^{\dagger} a \left(\eta b^\dagger - \eta^{*} b\right)} {\rm e}^{-{\rm i} b^{\dagger} b t'} \, ,
	\end{align}
	where $\eta=(1-{\rm e}^{-{\rm i} t'})$, $k=g / \omegaM$, $r= -\Delta / \omegaM$ and $t' = t\,\omegaM$ being the rescaled time in terms of the mechanical frequency $\omegaM$.  This time-evolution is obtained by applying the polaron transform used to obtain \eqnref{Eq:H_pol} and then transforming back into the standard frame.  Here, the non-linear nature of the dynamics is clear to see again with the Kerr-type term $(a^\dagger a)^2$ appearing in \eqnref{Eq:unitary_evolution_operator}.  Moreover, a periodic evolution is now apparent.  The third exponential is the only term responsible for dynamical evolution driven by $\eta$, which periodically returns to zero in $2 \pi$ (rescaled-)time intervals, i.e.~at the period of oscillation from the mechanical natural frequency.

	\subsection{Open dynamics}
	As indicated in \figref{Fig:Cavity1}, apart from accounting for laser driving, we model the decoherence by allowing the photons also to leave the cavity without being detected, and also couple the mechanical oscillator to a thermal bath that may spontaneously excite or destroy phonons. As a result, the reduced dynamics of the optomechanical system is described by the evolution of the density matrix incorporating mechanical and cavity degrees of freedom, $\rho$, through a master equation that reads~\cite{Breuer2002}:
	\begin{align} \label{Eq:master_full}
		\frac{\dd \rho}{\dd t} = & -\frac{\rm i}{\hbar} \left[\Hrf , \rho \right] + \kappa \left(a \rho a^\dagger - \frac{1}{2} \left[a^\dagger a , \rho \right]_+ \right) \nonumber \\
		& + \gamma \left(\bar{m} + 1 \right) \left(b \rho b^\dagger - \frac{1}{2} \left[b^\dagger b , \rho \right]_+ \right) \nonumber \\
		& + \gamma \bar{m} \left(b^\dagger \rho b - \frac{1}{2} \left[ b b^\dagger , \rho \right]_+ \right) \, ,
	\end{align}
 	where $\kappa$ is the (total) decay rate of the cavity mode, $\gamma$ is the damping rate of the mechanical mode and $\bar{m}=[\exp(\hbar\omegaM/\kB T)-1]^{-1}$ is the mean number of quanta in the thermal reservoir dictated by the mechanical frequency $\omegaM$ and the bath temperature $T$.  For small $\bar{m}$, this would require temperatures ranging from $10^{-8}$ -- $10^{-1}$K based on typical experimental numbers \cite{Aspelmeyer2014}.  As real-life experiments are typically conducted at cryogenic temperatures, this is a reasonable assumption to make in calculations.  Let us also remark that \eqnref{Eq:master_full} assumes that interactions with mechanics hardly affect the state of the bath, while the relaxation process of phonons occurs at timescales much larger than the actual mechanical dynamics ($\omegaM\gg\gamma\bar{m}$), so that Born-Markov and secular approximations safely apply~\cite{Breuer2002}.

	Although approximate solutions to dynamics \eref{Eq:master_full} are possible in restricted regimes~\cite{Mancini1997}, these should really be considered as `corrections' of the closed solution \eref{Eq:unitary_evolution_operator}, and are thus insufficient for our purposes. That is why, we resort to numerical methods in order to exactly solve \eqnref{Eq:master_full} by considering a truncated Fock space containing $\rho$, which we evolve then piecewise in time. In fact, such a solution describes then the \emph{ensemble average}, i.e.~the effective evolution of both the optical and mechanical models upon ignoring the records of a continuous photo-detection measurement, which we must now also incorporate into the model.

	\subsection{Unravelling the open dynamics}
	In order to facilitate numerical simulations, we take the approach of \emph{unravelling} the full dynamics \eref{Eq:master_full}, so that following the quantum-jump methodology~\cite{Dalibard1992,Molmer1993,Hegerfeldt1993,Carmichael1993,Plenio1998} and continuous measurement models \cite{Jacobs2014}, we first re-express \eqnref{Eq:master_full} as a non-linear stochastic equation preserving the purity, i.e.:
	\begin{widetext}
		\begin{align} \label{Eq:Full_stoch}
			\dd \rhoc = & -\frac{\rm i}{\hbar} \left[ \Hrf , \rhoc \right] \dd t - \kappa \left( \frac{1}{2} \left[ a^\dagger a , \rhoc \right]_+ - \Tr{a^\dagger a \rhoc} \rhoc \right) \dd t + \left(\frac{a \rhoc a^\dagger}{\Tr{a^\dagger a \rhoc}} - \rhoc \right) \dd N_t^{(\kappa)} \nonumber \\
			& - \gamma \left(\bar{m} + 1 \right) \left( \frac{1}{2} \left[ b^\dagger b , \rhoc \right]_+ - \Tr{b^\dagger b \rhoc} \rhoc \right) \dd t + \left(\frac{b \rhoc b^\dagger}{\Tr{b^\dagger b \rhoc}} - \rhoc \right) \dd N_t^{(\gamma_-)} \nonumber \\
			& - \gamma \bar{m} \left( \frac{1}{2} \left[ b b^\dagger , \rhoc \right]_+ - \Tr{b b^\dagger \rhoc} \rhoc \right) \dd t + \left(\frac{b^\dagger \rhoc b}{\Tr{b b^\dagger \rhoc}} - \rhoc \right) \dd N_t^{(\gamma_+)} \, ,
		\end{align}
	\end{widetext}
	where the $\dd N_t$-terms have the physical interpretation of random variables that represent counts over the infinitesimal time $\dd t$ of photon emission, phonon emission and phonon absorption, respectively, with expectation values~\cite{Jacobs2014}:
	\begin{align} \label{Eq:Full_weights}
		\left< \dd N_t^\kappa \right> & = \kappa\, \Tr{a^\dagger a \rhoc} \dd t \nonumber \\
		\left< \dd N_t^{\gamma_-} \right> & = \gamma \left(\bar{m} + 1 \right) \Tr{b^\dagger b \rhoc} \dd t \nonumber \\
		\left< \dd N_t^{\gamma_+} \right> & = \gamma \bar{m}\, \Tr{b b^\dagger \rhoc} \dd t \, .
	\end{align}
	The size of these expectation values gives then the respective weight of a jump of each form.  In the above and following, we use the subscript `c' to denote the dynamics of a \emph{conditional} state, i.e.~the evolution conditioned on a certain sequence of the jump statistics.  Formally, we write $\rhoc (t) \eqdef \rho (t|D_t)$, where $D_t = \{t_1 , t_2,\dots\}$ with $t_i \in \left[0,t\right]$ is the time at which the jump of a certain type occurs. 

	A quantum trajectory can then be generated using \eqnsref{Eq:Full_stoch}{Eq:Full_weights}.  This may be performed in an efficient manner, as rather than evolving stochastically \eqnref{Eq:Full_stoch} over each time-step $\dd t$, we may follow the quantum-jump prescription~\cite{Molmer1996,Plenio1998}. In particular, we may instead sample at what time the next jump happens, and propagate $\rhoc$ until that moment according to \eqnref{Eq:Full_stoch} conditioned on no jumps occurring. Only then, we determine the nature of the jump that occured from a ternary distribution with weights specified by the expectation values \eref{Eq:Full_weights} and apply the necessary jump operator. Such a procedure may then be repeated until the time $t$ of interest, while updating accordingly the expectation values \eref{Eq:Full_weights} with the current $\rhoc$ before implementing each jump.

	\subsection{Photon-counting as a continuous measurement}
	\label{sec:cont_photon_count}
	In reality, however, we have access only to data $D_t$ of quantum jumps caused by the detected photons. Hence, to obtain the observed conditional dynamics we must average \eqnref{Eq:Full_stoch} over stochastic jumps occurring due to emission of unobserved photons and both excitations and emission of phonons.  Still, by averaging the full stochastic dynamics \eref{Eq:Full_stoch} over the inaccessible degrees of freedom we arrive at the dissipative conditional evolution of the optomechanical system:
	\begin{widetext}
		\begin{align} \label{Eq:Detector_stoch}
			\dd \rhoc = & -\frac{\rm i}{\hbar} \left[ \Hrf , \rhoc \right] \dd t + \left(\kappal a \rhoc a^\dagger - \frac{\kappa}{2} \left[ a^\dagger a , \rhoc \right]_+ + \kappad \,\Tr{a^\dagger a \rhoc} \rhoc \right) \dd t \nonumber \\
			& + \gamma \left(\bar{m}+ 1 \right) \left( b \rhoc b^\dagger - \frac{1}{2} \left[b^\dagger b , \rhoc \right]_+ \right) \dd t + \gamma \bar{m} \left(b^\dagger \rhoc b - \frac{1}{2} \left[b b^\dagger , \rhoc\right]_+ \right) \dd t \nonumber \\
			& + \left(\frac{a \rhoc a^\dagger}{\Tr{a^\dagger a \rhoc}} - \rhoc \right) \dd N_t^{\kappad},
		\end{align}
	\end{widetext}
	where the \emph{true} conditional state $\rhoc$ describing cavity and mechanical modes no longer maintains purity, and only the decay rate associated with detected photons, $\kappad$, parametrises now the stochastic quantum jumps with $\left< \dd N_t^\kappad \right> = \kappad \Tr{a^\dagger a \rhoc} \dd t$ at a given time $t$.
	
	In what immediately follows however, we initially consider the simplified situation with all the sources of decoherence being absent, i.e.~$\kappal = \gamma = 0$ and only $\kappad > 0$ in \eqnref{Eq:Detector_stoch}, to observe the impact of conditioning on photon-detection.  In such a simple case, the dynamics follows two types of evolution: unitary dynamics under `\emph{no-photon}' detection (denoted `\emph{no-ph}' for short) and a quantum jump whenever a detection occurs.  The former, being formally a special case of the conditional dynamics \eref{Eq:Detector_stoch} with $D_t = \emptyset$, can be naturally modelled by ignoring the normalisation of $\rhoc$ and applying to it the time-evolution operator $U_\nph(t)=\exp\{-\ii H_\nph\,t/\hbar\}$ generated by the non-Hermitian Hamiltonian:
	\begin{equation} \label{Eq:H_cond}
		H_{\nph}=\Hrf - \frac{\ii \hbar \kappad}{2} a^{\dagger} a \,,
	\end{equation}
	which is obtained by gathering the (non-zero) Hamiltonian-like terms in \eqnref{Eq:Detector_stoch}.

	In such a picture, given that the optomechanical system is initialised in a pure state $\rhoc(0)=\proj{\psic(0)}$, the probability of not observing any emission until time $t$ is just $P_\nph(t) = || \ket{\psic(t)} ||^2$, where $\ket{\psic(t)}= U_\nph(t) \ket{\psic(0)}$ is the unnormalised conditional state that must then be updated according to $\ket{\psic(t)} \rightarrow a \ket{\psic(t)}$ whenever a jump occurs, renormalised, and evolved again by $U_\nph(t)$ until the next jump.  Moreover, in the regime of no driving the time-evolution operator can be explicitly written in a form similar to \eqnref{Eq:unitary_evolution_operator} as
	\begin{align}
		\label{Eq:Non_unitary_evolution_operator}
		U_\nph(t') = & {\rm e}^{-{\rm i} \tilde{r} a^{\dagger} a t'} {\rm e}^{{\rm i} k^2 (a^{\dagger} a)^2 (t' -\sin (t'))} \nonumber \\
		& \times {\rm e}^{k a^{\dagger} a (\eta b^\dagger - \eta^* b)} {\rm e}^{-{\rm i} b^{\dagger} b t'} \, ,
	\end{align}
	where the only difference from \eqnref{Eq:unitary_evolution_operator} is that now $r\to\tilde{r}$ with $\tilde{r} = - \left(\Delta + {\rm i} \kappad / 2 \right) / \omegaM$.

	\section{Enhancing entanglement by photon-counting}
	\label{sec:entanglement}
	As continuous monitoring clearly modifies the evolution of an optomechanical system, as can be seen from comparing \eqnref{Eq:master_full} with \eqnref{Eq:Detector_stoch}, we first investigate how it affects the dynamics of entanglement between the cavity and mechanical modes, before moving towards sensing applications.  We start by assuming that neither driving nor decoherence is present in the system, in order to then consider the complete dissipative dynamics.

	\subsection{Closed conditional dynamics}

	In absence of driving and unobserved dissipation, we may just use the (non-unitary) time-evolution operator \eref{Eq:Non_unitary_evolution_operator} to propagate the optomechanical state in between the jumps. Furthermore, as decoherence is absent, the state preserves its purity throughout the evolution. Hence, after choosing a pure initial state we may quantify the entanglement between the cavity and mechanical modes at any time through the \emph{linear entropy} (of the reduced state)~\cite{horodecki_quantum_2009}, i.e.:
	\begin{equation}
		\label{Eq:Entropy}
		S(t) \eqdef 1 - P(t) \,,
	\end{equation}
	where by $P\eqdef\Tr{\cav{\rho}^2}$ we denote the purity of the cavity's reduced state $\cav{\rho}=\mech{\trace}\{\rho\}$, with $\rho$ being the density matrix of the total system. Moreover, we assume here both the cavity and the mechanics to be prepared in coherent states, i.e.~$\ket{\psic(0)} = \cav{\ket{\alpha}} \mech{\ket{\beta}}$ with complex amplitudes $\alpha$ and $\beta$, respectively, in which case the evolution of purity $P(t)$ under the conditional `no-photon' evolution \eref{Eq:Non_unitary_evolution_operator} can be evaluated explicitly, see \appref{ap:app1}.  

	\begin{figure}[t!] 
		\centering
		\includegraphics[width=0.99\linewidth]{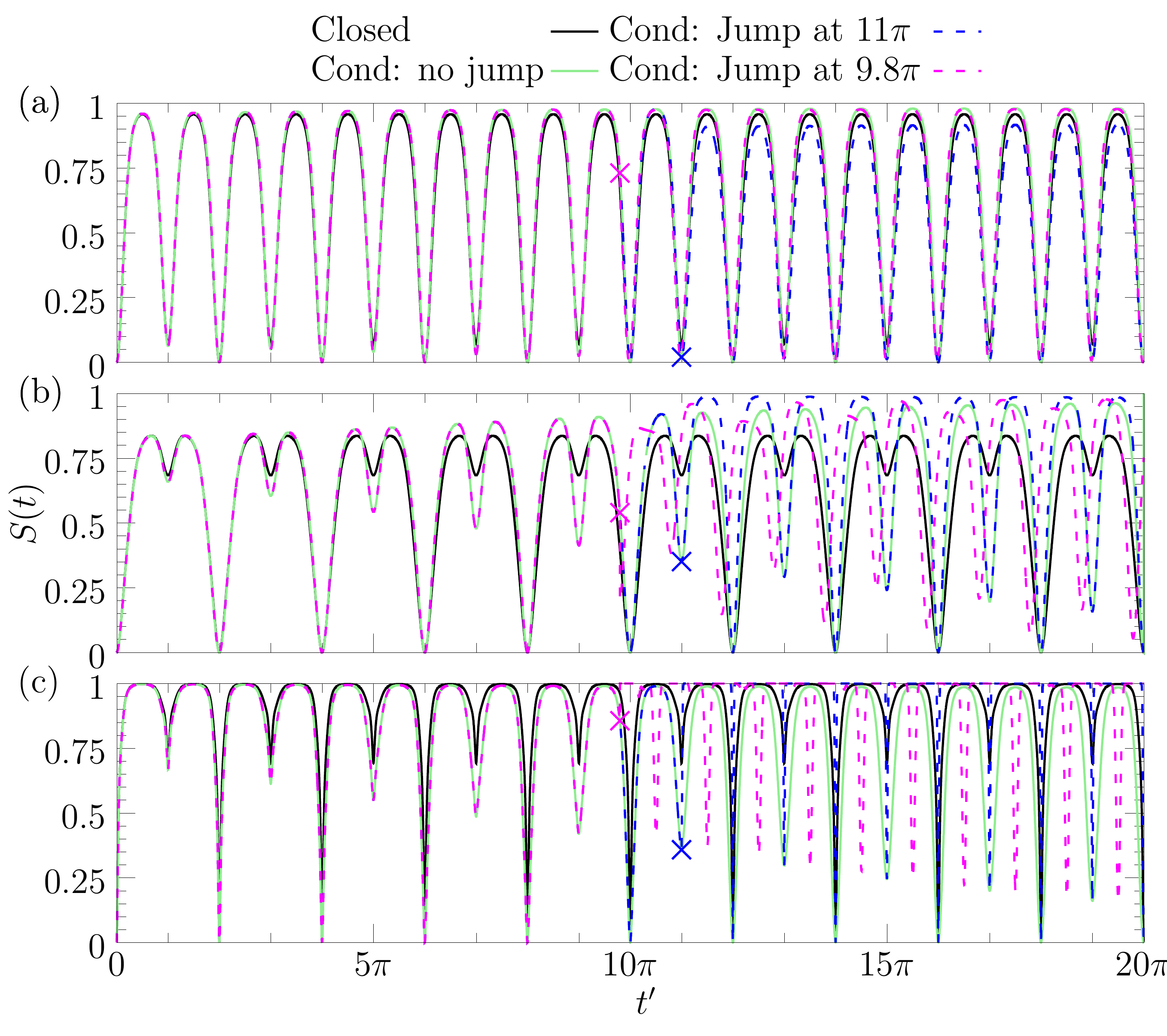}
		\caption{ 
			\textbf{Entanglement dynamics of a closed system and the impact of a quantum jump}, illustrated through the linear entropy $S$ of the cavity's reduced state.	The system is initialised in coherent states of optics and mechanics, $\cav{\ket{\alpha}} \mech{\ket{\beta}}$ with $\alpha = \beta = 1$, with the optomechanical coupling set to $g = \{0.1,1,5\}$ in plots (a), (b) and (c), respectively, for $\omegaM = 1$.  For the conditional dynamics, we set the cavity decay rate at which the photons are detected to $\kappad=0.04$.  The closed system (\emph{solid black line}) periodically returns to its initial, unentangled state at intervals of $2 \pi$.  When loss through photon-counting is added, the conditional state in the absence of any jumps produces similar entanglement characteristics, with a slight modulation of the oscillations due to the exponential decay of the cavity's amplitude (\emph{solid light-green line}).  After a jump the behaviour of the linear entropy deviates from the conditional `no-photon' evolution, depending on when the jump happens, as seen in the \emph{dashed pink and blue lines} for the jumps occurring at $9.8\pi$ and $11\pi$ (\emph{crosses}), respectively.
		}
		\label{Fig:Entanglement_closed}
	\end{figure}
	
	In \figref{Fig:Entanglement_closed}, we present how the linear entropy evolves over time for the three different cases of:~%
	closed dynamics (in absence of photon-detection), conditional dynamics with photon-detection but no photon-clicks observed, and conditional dynamics with a single jump occurring at some time. As the closed evolution is periodic, see \eqnref{Eq:unitary_evolution_operator}, we consistently observe that the entanglement is continuously created and destroyed, while the system returns back to its initial state at any $t' = 2 \pi \ell$ with $\ell\in\mathbb{N}$. In case of conditional `no-photon' dynamics \eref{Eq:Non_unitary_evolution_operator} similar behaviour is observed but the system returns rather to $\cav{\ket{\exp(-\kappad t'/2\omegaM)\alpha}} \mech{\ket{\beta}}$ at any $t' = 2 \pi \ell$. Hence, the evolution of linear entropy is no longer exactly periodic with its oscillations growing due smaller and smaller amplitude of the cavity mode at the start of each period.

	Unless the jump occurs exactly at $t' = 2 \pi \ell$, at which the modes are not entangled and the cavity is in a coherent state, it affects the state and the subsequent evolution changes -- see dashed lines in \figref{Fig:Entanglement_closed}.  This can be intuitively understood in the Fock basis of the cavity, as can be seen in \appref{ap:app1}:~the quantum jump shifts then by one the corresponding components, such that each Fock state of the cavity does not couple to the same mechanical state as before the jump.  Consequently, the original pairs of cavity and mechanical states start to evolve out of phase with one another, and the system no longer returns periodically to a product state. One corollary of this is that some level of entanglement is maintained at all times after a jump, as can be seen from \figref{Fig:Entanglement_closed}.
	
	\begin{figure*}[t] 
		\centering
		\includegraphics[width=0.99\linewidth]{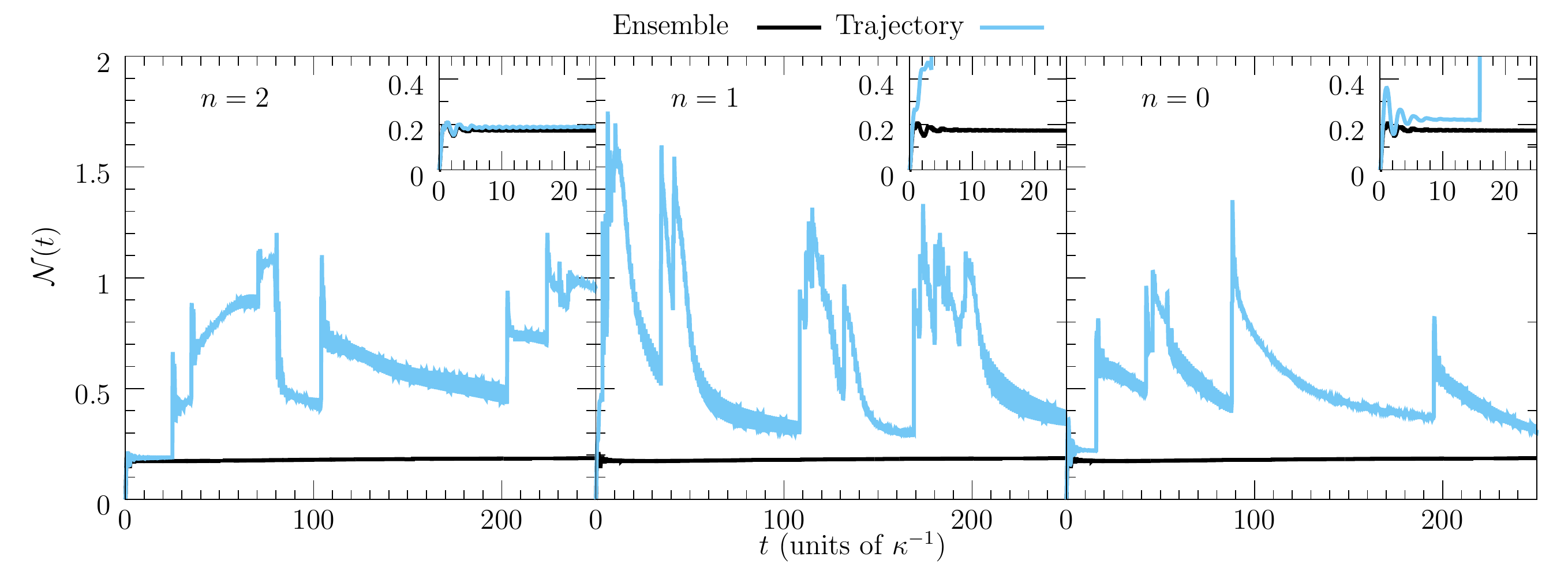}
		\caption{ 
			\textbf{Entanglement dynamics of an open system for the three non-linear regimes of interest}, illustrated through negativity $\mathcal{N}(t)$ evaluated between the truncated cavity and mechanical modes.  Red-detuning of the driving field is set to  $\Delta = - n \,g^2 / \omegaM$ for (from \emph{right to left}) the on-resonance ($n=0$), blockade ($n=1$) and cascade ($n=2$) non-linear regimes. The evolution of negativity is depicted for the ensemble \eref{Eq:master_full} (\emph{black}) and conditional \eref{Eq:Detector_stoch} dynamics along an exemplary photon-click trajectory (\emph{blue}) generated for each case.  Each photon-detection significantly boosts the negativity, before it subsequently decays towards the stationary value of the ensemble.  Although the general behaviour of negativity is similar in all three cases,  the exemplary photon-click patterns already indicate that each regime exhibits different (non-classical) photon statistics~\cite{Kronwald2013}.  We take parameters of $g/4 = \omegaM/4\sqrt{2} = \kappa$, with $\kappad = 9 \kappal$, while $\Omega = 0.3 /\omegaM$, $\gamma = 10^{-3} \omegaM$ and $\bar{m}= 1$.  The detunings taken correspond to the regimes discussed in the main text.
		}
		\label{Fig:Entanglement_open}
	\end{figure*}

	\subsection{Open conditional dynamics}
	\label{sec:Open_dynamics}
	We move onto the main focus of our work that is the realistic scenario with open dynamics \eref{Eq:Detector_stoch}, which includes extra unobserved loss of photons ($\kappal>0$) and damping of the mechanical oscillator via the phonon-exchange with a thermal reservoir ($\gamma,\bar{m}>0$). Moreover, as the investigation of non-linear effects is our priority, we consider the previously mentioned three scenarios of (red-detuned) driving with $\Delta = - n g^2 / \omegaM$ and $n=\{0,1,2\}$, so that in each of these cases the multi-photon transition $\cav{\ket{0}} \rightarrow \cav{\ket{n}}$ is favoured -- see \eqnref{Eq:energy_levels}. In particular, similarly to \citeref{Kronwald2013}, we set $g/4 = \omegaM / 4 \sqrt{2} = \kappa$ and the driving strength to $\Omega = 0.3 / \omegaM$, and refer to three cases as \emph{on-resonance} ($n=0$), \emph{blockade} ($n=1$) and \emph{cascade} ($n=2$) regimes, even though to actually observe ``blocking'' of photons (i.e.~their explicit antibunching at short time-scales) for $n=1$, we would need to decrease $\Omega$~\cite{Kronwald2013}. We also set the mean number of excitations in the mechanical reservoir to $\bar{m} = 1$, and the decay-rates to $\gamma = 10^{-3} \omegaM$, $\kappad = 0.9 \kappa$ and $\kappal = 0.1 \kappa$, i.e.~accounting also for imperfect detection with 10\% of photons being effectively lost.  This parameter choice satisfies the non-linearity condition of Eq.~(\ref{Eq:Non-lin}), but also yields the so-called \emph{sideband-resolved regime} of the cavity~\cite{Aspelmeyer2014}, in which the emitted photons could in principle be further classified by measuring their frequency. However, the frequency resolution is unnecessary in such a setting to observe non-classical statistics of photon detections~\cite{Kronwald2013}, as these emerge due to the non-linearity that is the key resource utilised here for sensing applications.  We further show this within \appref{app:sensing_simple}, where we consider dynamics outside the sideband-resolved regime and are still able to perform sensing based on the emergent photon-click trajectories. Throughout the rest of the paper, the above parameter values shall be taken, unless otherwise stated.

	Being in contact with a thermal reservoir, the initial state of the mechanical oscillator must be in a thermal state of the form~\cite{ferraro_gaussian_2005}:
	\begin{align}
		\mech{\rho}^{\rm th} = & \left(1 - \frac{\bar{m}}{\bar{m}+1} \right) \sum_{m=0}^\infty \left(\frac{\bar{m}}{\bar{m}+1}\right)^m \ket{m} \bra{m} \, ,
	\end{align}
	while we assume the cavity to be initially empty. Hence, the optomechanical system is always initialised in
	\begin{equation}
		\rho(0) = \cav{\ket{0}}\!\bra{0} \otimes \mech{\rho}^{\rm th} \,,
	\end{equation}
	and the source of photons that enter the cavity and then leak to be detected is consistently the driving field.
	
	Dealing with open-system dynamics, we consider then the \emph{negativity} as a valid quantity that witnesses entanglement, being defined as~\cite{guhne_entanglement_2009,horodecki_quantum_2009}:
	\begin{equation}
		\mathcal{N}(t) \eqdef \frac{\left\| \rho(t)^{\Gamma_{\rm M}}\right\|_1 - 1}{2} \, ,
	\end{equation}
	where by $\left\|A\right\|_1\eqdef\Tr\sqrt{A^\dagger A}$ we denote the trace-norm of a matrix $A$, and $\rho^{\Gamma_{\rm M}}$ corresponds to taking the partial transpose over the (truncated) subspace associated with the mechanical mode.

	In \figref{Fig:Entanglement_open}, we present the evolution of negativity for three exemplary photon-click trajectories (\emph{blue lines}), generated in the non-linear regimes of interest, determined by adjusting the (red) detuning with $n=\{0,1,2\}$. We observe that even for ensemble dynamics (\emph{black lines}) the entanglement is witnessed with negativity saturating at a similar non-zero value in all three regimes, while the system reaches its stationary state that must be entangled.  This immediately shows the impact of non-linear interactions being present, as within the linearised regime the red-detuned dynamics would lead to an effective beamsplitter interaction between the optical and mechanical modes, and hence a negligible amount of entanglement being acquired between them~\cite{Aspelmeyer2014}.  Contrastingly, the non-linear interaction here allows for the generation of non-classical stationary states.

	However, in each case the detection of a photon sharply increases $\mathcal{N}$, which afterwards exponentially drops towards some constant value, unless another click occurs. This `steady' value for the conditional dynamics is higher than for the ensemble  due to an effective reduction in noise -- with $\kappad$ no longer contributing to decoherence, while other sources of dissipation remain. Finally, even though only one trajectory is presented for each of the non-linear regimes, by inspecting the peaks in negativity it is evident that photon-detection statistics are very different in the on-resonance ($n=0$), blockade ($n=1$) and cascade ($n=2$) regimes, with the latter two exhibiting apparent bunching of emitted photons~\cite{Kronwald2013}. We now study these non-classical effects in detail for each case.

	\section{Statistics of detected photons}
	\label{sec:photon_statisics}
	In order to investigate the statistics of detected photons for the three detuning regimes, we firstly sample over many numerically generated quantum trajectories to determine the relative likelihood, $\zeta$, of the next photon emission at a time $t_2$ after a photon emission at time $t_1 \leq t_2$ defined as
	\begin{align}
		\zeta \eqdef & \frac{N_{t_1,t_2}}{N_{\rm tot}} \, ,
	\end{align}
	with $N_{t_1,t_2}$ being the number of sampled photons observed at $t_2$ after an emission at $t_1$, while $N_{\rm tot}$ being the total number of sampled photons.  We observe that the cascade regime ($n=2$) favours an immediate photon emission after a first photon, especially when compared with the blockade regime ($n=1$), thus demonstrating stronger photon bunching. 

\begin{figure}[t!] 
		\centering
		\includegraphics[width=0.99\linewidth]{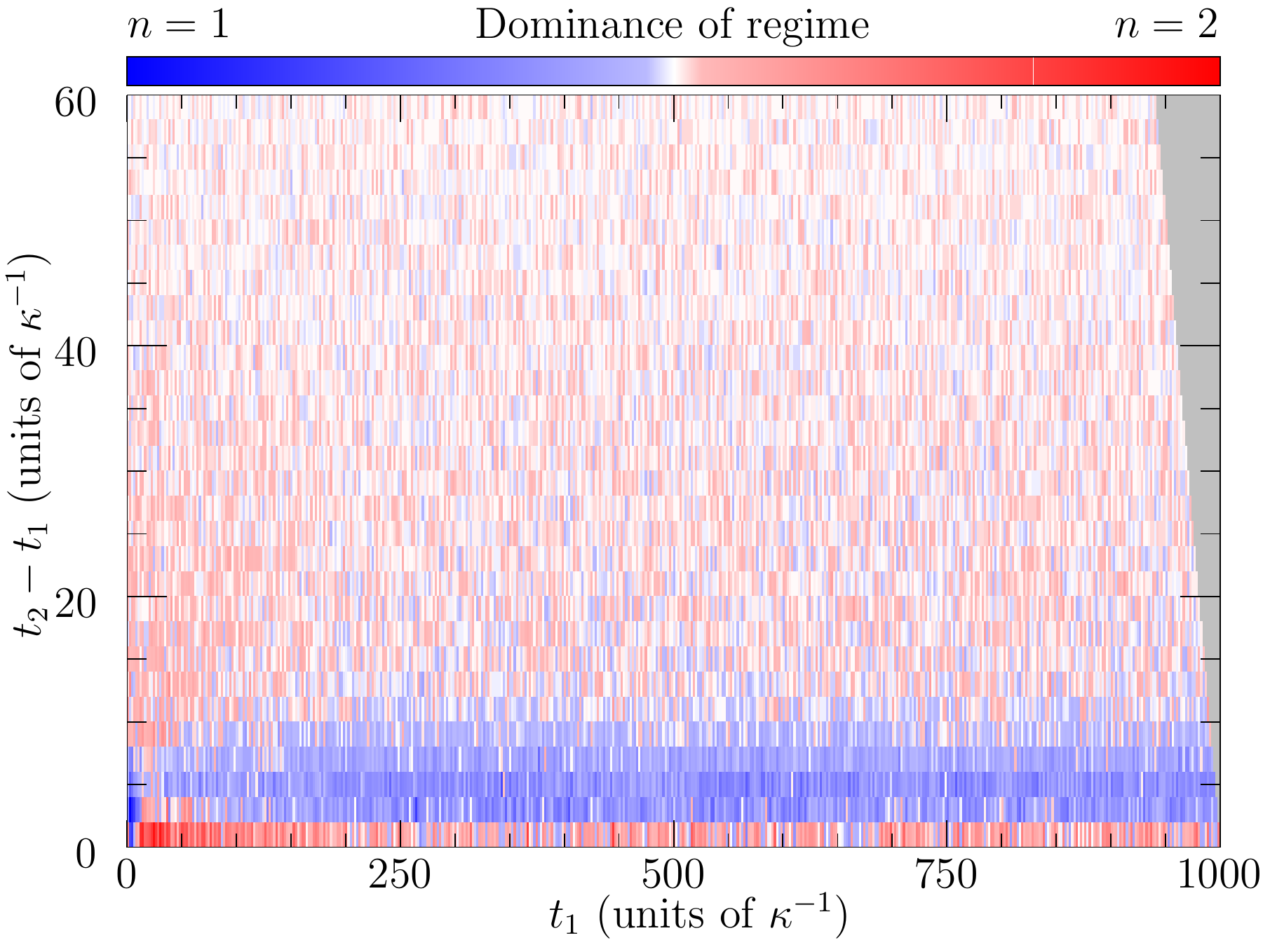}
		\caption{ 
		\textbf{Blockade or cascade regime dominating the photon-counts for a given time separation.}
			The colour of each block, and its intensity, shows whether the particular time delay between two photon-clicks is favoured by either the blockade (\emph{blue}, $n=1$) or cascade (\emph{red}, $n=2$) regime, with white entries indicating no significant difference.   The above data is obtained by sampling over 1000 trajectories for each detuning regime, while the grey triangle on the right edge appears due to no data available for these late entries.  For small time differences $\Delta t\eqdef t_2 - t_1\le2$, an emission in the ($n=2$)-regime is generally more likely, due to the photon cascade behaviour (narrow red horizontal patch). In contrast, it is the blockade regime ($n=1$) in which more photons pairs are on average emitted with $2\le \Delta t\le 10$ (wide blue horizontal patch). From the horizontal variation of the figure, one can infer that around $t_1\approx100$ the optomechanical system reaches its stationary-state behaviour with the strength of two-photon correlations depending then only on $\Delta t$ (similar vertical pattern).
		}
		\label{Fig:Samp}
	\end{figure}

	In order to show this non-classical effect explicitly, we plot in \figref{Fig:Samp} the difference of the $\zeta$ values for $n=2$ and $n=1$ for an array of time bins over $t_1$ and $t_2$.  At any timescale (value of $t_1$), it is clear that for small time-delays, $\Delta t\eqdef t_2 - t_1$, photon emissions are observed more frequently in the $n=2$ regime (narrow red horizontal patch for $\Delta t\le 2$), whereas for slightly larger delays observing emissions in the $n=1$ regime becomes more likely (wide blue horizontal patch for $2\le \Delta t\le 10$).  As expected, for large differences between emission times, no strong correlations between subsequent photons are visible with both regimes ($n=2$ and $n=1$) yielding similar numbers of detected photons (no dominating colour for $\Delta t\ge 10$).

	Furthermore, it is also apparent from \figref{Fig:Samp} that the two-photon correlations strongly depend on the time of observing the first emission (value of $t_1$), as we are dealing with dynamics of a quantum system requiring some time to reach a stationary state (around $t_1\approx100$ in \figref{Fig:Samp}, from which the overall pattern stabilises). One should bear in mind this fact when computing the second-order correlation, $g^{(2)} (t_1 , t_2)$ depending not only on $\Delta t$, as we now demonstrate.

	\begin{figure*}[t] 
		\centering
		\includegraphics[width=0.99\linewidth]{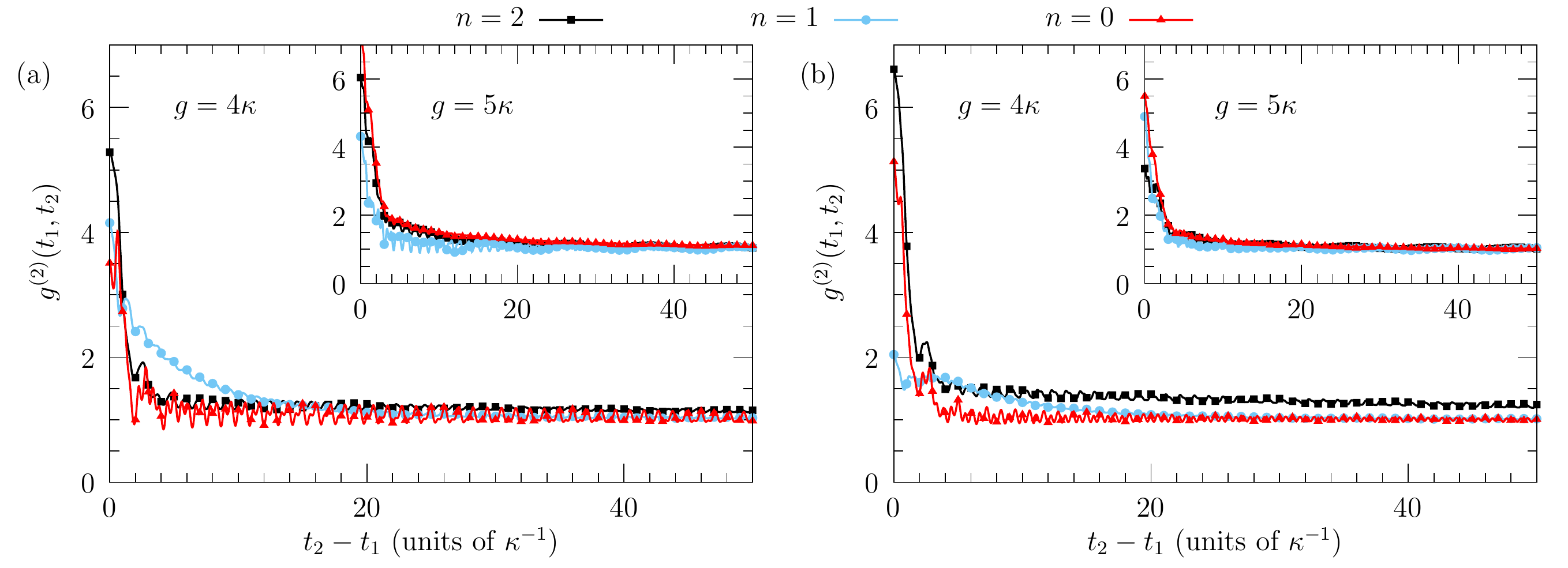}
		\caption{
			\textbf{Comparison of $g^{(2)}$-functions  within each detuning regime.}  We plot for the parameters described in \secref{sec:Open_dynamics}, both at (a) an early time in the evolution and (b) later within the stationary regime.  At both times, we see similar behaviour, where the $n=2$ detuning shows the strongest bunching.  Within the stationary regime, the $n=1$ detuning has a peak a short time after the emission of a photon, rather than immediately after in the $n=2$ case, showing that this regime favours a gap in emissions more than the $n=2$ case.  For $n=0$, although there is an initial peak, this decay rapidly towards no correlations.  Within the inset plots, we plot for $g=5$ instead to see how the statistics change with $g$, while keeping the value for the detuning the same, i.e.~as for $g=4$.  Here, $n=2$ no longer has the largest initial peak, but all correlations rapidly decay.  The remaining parameters are set as in \figref{Fig:Entanglement_open}.
		}
		\label{Fig:g2_g4}
	\end{figure*}

	In particular, we define the second-order correlation function for the photon-clicks as
	\begin{align}
		\label{Eq:g2}
		g^{(2)} (t_1 , t_2) \eqdef & \frac{p (t_1 , t_2)}{p(t_1) p(t_2)} = \frac{p(t_2 | t_1)}{p(t_2)} \, ,
	\end{align}
	being determined by the probability of detecting a photon at the respective time arguments, where we assume $t_2 \geq t_1$. However, rather than resorting to sampling over trajectories, we now utilise the fact that the $g^{(2)}$-function can be equivalently computed by evolving the system according to the ensemble-averaged dynamics \eref{Eq:master_full} before and in between the two emission-points of interest~\cite{Gardiner2004}. Writing the reset superoperator as $\mathcal{A}[\cdot] \eqdef a [\cdot] a^\dagger$ and denoting the time-evolution superoperator from $t_1$ to $t_2$ as $\mathcal{T}_{t_2 , t_1}$~\cite{Gardiner2004}, i.e.~the solution to the master equation \eref{Eq:master_full}, the $g^{(2)}$-function reads
	\begin{align}
		g^{(2)} (t_1 , t_2) = & \frac{\trace \left( \mathcal{A} \mathcal{T}_{t_2 , t_1} \mathcal{A} \mathcal{T}_{t_1 , t_0} \rho(t_0) \right)}{\trace \left(\mathcal{A} \mathcal{T}_{t_1 , t_0} \rho (t_0)\right) \trace \left(\mathcal{A} \mathcal{T}_{t_2 , t_0} \rho (t_0)\right)} \, .
	\end{align}
	
	The above expectation values can be extracted from numerical solutions to the master equation~\eref{Eq:master_full}, the results of which are shown in \figref{Fig:g2_g4}.  Here, we consider two different intial times $t_1$ to reflect the differences in the photon statistics at short and long times.  This is because we initialise the cavity state in the vacuum, whereas at long times it is in its stationary state. In both cases, we see an initially very large value for the $n=2$ detuning, demonstrating its photon-cascade behaviour, whereas we see a lower value for the $n=1$ case, demonstrating some blockade effects.  We note that we do not see true blockade statistics here with $g^{(2)} < 1$ due to the overall strength of the driving. 
	In order to observe explicit antibunching of photons	a weaker driving would be required to inhibit
	the population of higher Fock states~\cite{Kronwald2013}. Nevertheless, the bunching is suppressed compared with the $n=2$ case. 
	
	Finally, as we are ultimately interested in using these photon statistics for sensing, we consider how they are altered by changing another system parameter that is to be inferred, while keeping all other parameters the same.  In this study, we focus on estimating the optomechanical coupling strength $g$.  From \figref{Fig:g2_g4} we see that the photon statistics clearly change for a different value of $g$.  This is due to the detuning no longer being tailored to the correct value of $g$, inhibiting the blockade/cascade effects, as there is no longer balance in Eq.~(\ref{Eq:energy_levels}) creating energetically favourable transitions.  In particular, we see that the enhanced bunching for the $n=2$ case is lost in the stationary period -- compare black lines in Fig.~\ref{Fig:g2_g4}(b), especially at $\Delta t=0$;~while for $n=1$ the enhancement in subsequent emission is suppressed in the early-times regime -- compare the drop of blue lines in Fig.~\ref{Fig:g2_g4}(a) at small $\Delta t$. Crucially, such observations confirm that a shift in the value of a parameter has a noticeable effect on the photon statistics, demonstrating that the photon-click pattern contains information suitable 
	for parameter inference.

	\section{Methods for quantum sensing in real time}
	\label{sec:bayesian}

	Now that we have an understanding of the physical behaviour of the system and the photon statistics that it generates, we may consider how it can be used for sensing applications.  Rather than perform a single measurement on the system at a certain time, we wish to continuously gain information about the system by monitoring its photon statistics.  In order to do so, we resort to Bayes's rule, which states
	\begin{align} \label{Eq:Bayes}
		P(\theta | D_t) = & \frac{P(D_t | \theta) P(\theta)}{P(D_t)} = \frac{P(D_t | \theta) P(\theta)}{\int \dd \theta P(D_t | \theta) P(\theta) } \, ,
	\end{align}
	for a parameter $\theta$ to be inferred from some click-pattern, $D_t$, up to time $t$.  The data we use here is the photon emission statistics of the optical cavity observed in the detector channel (with rate $\kappad$), while we consider the parameter of interest to be 
	$g$, although any system parameter affecting $D_t$ could be chosen and inferred in this way.
	
	\subsection{Evaluating probabilities of particular trajectories}
	In this subsection, we shall describe how the probability $P(D_t|\theta)$ can be calculated numerically.  We use the conditional Hamiltonian \eref{Eq:H_cond} to describe the evolution under no detections,	while simultaneously accounting for decoherence due to loss of undetected photons and dissipation of phonons by resorting to the standard Lindbladian description, as in \eqnref{Eq:Detector_stoch}.  The result is a conditional master equation describing the dynamics of an unnormalised density matrix, $\sigma_\nph$, i.e.
	\begin{align} \label{Eq:Master_cond}
		\frac{\dd {\sigma}_\nph}{\dd t} = & -\frac{\rm i}{\hbar} \left(H_\nph \sigma_\nph - \sigma_\nph H_\nph^\dagger  \right) \nonumber \\
		& \hspace{-0.8cm} + \kappa_{\rm l} \left(a \sigma_\nph a^\dagger - \frac{1}{2} \left[a^\dagger a , \sigma_\nph\right]_+ \right) \nonumber \\
		& \hspace{-0.8cm} + \gamma \left(\bar{m} + 1 \right) \left(b \sigma_\nph b^\dagger - \frac{1}{2} \left[b^\dagger b , \sigma_\nph \right]_+ \right) \nonumber \\
		& \hspace{-0.8cm} + \gamma \bar{m} \left(b^\dagger \sigma_\nph b - \frac{1}{2} \left[b b^\dagger , \sigma_\nph\right]_+ \right) \, .
	\end{align}
	This equation can be solved numerically in order to obtain the conditional density matrix under no detected photons at a given time $t_1$. Adapting the notation of the previous section, we write
	\begin{align} \label{Eq:no-ph_evolution}
		\sigma_\nph (t_1) = & \, \mathcal{T}_{t_1 , t_0}^\nph \rho (t_0) \, ,
	\end{align}
	assuming a fully normalised density matrix at time $t_0$.  This describes the conditional (no-photon) evolution of the density matrix from $t_0$ to $t_1$.  Just as in the closed system case that we discussed, the trace of this density matrix is not preserved due to the effect of $H_\nph$, with its value corresponding to the `no-photon' probability, i.e.:
	\begin{equation} \label{Eq:P0}
		P_\nph (t_1 , t_0) = \Tr{\mathcal{T}_{t_1 , t_0}^\nph \rho (t_0)} \, ,
	\end{equation}
	which is the probability of detecting no photons between $t_0$ and $t_1$, i.e.~the \emph{exclusive} probability~\cite{Plenio1998}. As a consequence, one may define the probability density $I_1(t_1,t_0)$ for the emission of a single photon just after the time $t_1$ as~\cite{Plenio1998}:
	\begin{align} \label{Eq:w1}
		 I_1(t_1 , t_0) \eqdef & - \frac\dd{\dd t_1} P_\nph (t_1, t_0) \, .
	\end{align}
	Using \eqnsref{Eq:P0}{Eq:w1}, and a discretisation of time steps of size $\Delta t$, where $\Delta t \ll \kappa_\dd$ such that the probability of multiple photon emissions within the time-bin $\Delta t$ is negligible, we can obtain the probability $P(D_t|\theta)$.  Given there are $N$ photon-click events within total time $t$ of $D_t$, i.e.~$|D_t| = N$, the desired conditional probability reads
	~\cite{Plenio1998}:
	\begin{align}
		\label{eq:click_decomp}
		P(D_t|\theta) = & P_\nph(t, t_N) \prod\limits_{n=1}^N \Bigg( P_\nph (t_n - \Delta t , t_{n-1}) \nonumber \\
		& \times \int\limits_{\Delta t}^{0} \dd t' I_1(t_n , t_n - t') \Bigg) \, ,
	\end{align}
	where after each photon-click event at $t_n$, the density matrix is 
	reset as follows, $\rho(t_n) \rightarrow a \rho_0 (t_n) a^\dagger / \trace \left(a^\dagger a \rho_0 (t_n)\right)$, leading to the evolution of an initially normalised density matrix as in \eqnref{Eq:no-ph_evolution}.  In this way, each probability element $P_\nph$ and $I_1$ in \eqnref{eq:click_decomp} is evaluated using the relevant initial normalised state $\rho(t_n)$.

	As the conditional probability $P(D_t | \theta)$ is defined for a particular value of the parameter $\theta$, in order to obtain the relevant posterior distribution $P(\theta| D_t)$ in \eqnref{Eq:Bayes},	we must evaluate $P(D_t | \theta)$ for many different values of $\theta$. The distribution $P(D_t | \theta)$ is obtained by numerically determining the probability of observing a given photon-click trajectory $D_t$ for a sufficiently dense grid of $\theta$-values. This means we have access to the full probability distribution, rather than relying on sampling techniques such as the Metropolis-Hastings algorithm~\cite{Gammelmark2013}.  However, we observe that the improved methods of sampling, also e.g.~ones employing so-called particle filters, may benefit significantly the computation speed of our Bayesian inference procedure, only if we extended our analysis to problems of sensing multiple (or multidimensional) parameters \cite{Murphy2012}, while the grid-based approach remains efficient in single-parameter sensing. In fact, the computation bottleneck is dictated by the process of evaluating the likelihood \eref{eq:click_decomp} for any given trajectory $D_t$.

	\subsection{Bayesian inference}
	\label{subsec:bayinfer}
	One of the key features of Bayesian inference is that it provides solution that depend on the \emph{prior distribution}~\cite{Kay1993}, which must be selected to most accurately represent our \emph{a priori} knowledge about the problem -- here the parameter $\theta$ to be inferred. In this work we choose the prior to read~\cite{Li2018}:
	\begin{align} \label{Eq:Prior}
		P(\theta) = & \frac{1}{\theta_{\rm max} - \theta_{\rm min}} \frac{\exp \left(\alpha \sin^2 \left(\frac{\pi \left(\theta - \theta_{\rm min}\right)}{\theta_{\rm max} - \theta_{\rm min}}\right) \right) - 1}{\exp \left(\frac{\alpha}{2} \right) I_0 \left(\frac{\alpha}{2}\right) - 1} \, ,
	\end{align}
	so that by choosing $\theta_{\rm max/min}$ we can put strict constraints on the range of parameters to be considered.  Moreover, $I_0$ is a zeroth order modified Bessel function of the first kind, while $\alpha$ is a real number that controls the sharpness of the distribution, which gets flatter as $\alpha$ becomes more negative. Let us also note that this distribution is both continuous and zero at its boundaries, and hence allows for application of Van Trees bounds on the average estimation performance~\cite{van2007bayesian}, which we discuss below.

	One of the key parts of any useful sensing scheme is the efficient estimator, whose role is to most accurately provide the value of the unknown parameter based on the data available. It is common to consider the mean of the posterior distribution as the \emph{optimal} estimator for Bayesian inference, $\estB{\theta}(D_t) \eqdef \int\!\dd \theta\, P(\theta|D_t)\,\theta$, as it minimises on average the \emph{mean-squared error} (MSE) for the true value of the parameter, $\theta$, being drawn from the prior~\cite{Kay1993,Trees1968},
	\begin{equation} \label{Eq:Av_MSE}
		\av{\Delta^2 \est{\theta}}{P(\theta)} \eqdef \int\!\dd \theta\,P(\theta)\,\Delta^2 \est{\theta}|_\theta,
	\end{equation} 
	whose minimum corresponds then to the variance of the posterior distribution, $P(\theta|D_t)$, averaged over all data patterns, i.e.:
	\begin{equation}
		\av{\Delta^2 \estB{\theta}}{P(\theta)} = \int\!\dd D_t\,P(D_t)\;\Var{\theta}{P(\theta|D_t)}.
	\end{equation} 
	In \eqnref{Eq:Av_MSE}, the MSE for a particular value of $\theta$ is generally defined and may be decomposed as:
	\begin{align} \label{Eq:MSE}
		\Delta^2 \est{\theta}|_\theta 
		& \eqdef  \int\!\dd D_t\, P(D_t|\theta)\,\left(\est{\theta}(D_t)-\theta\right)^2 \\
		& =  \Var{\est{\theta}}{P(D_t |\theta)} + \left(\av{\est{\theta}}{P(D_t|\theta)} - \theta\right)^2 \, , \nonumber 
	\end{align}
	where the first and second terms above represent the uncertainty and bias of any estimator, respectively.  Note that when considering a finite dataset (e.g.~a single photon-click trajectory of finite time duration) the bias may not be ignored, as any well-behaved estimator, e.g.~$\estB{\theta}$, is guaranteed to become \emph{unbiased} only in the asymptotic limit of many independent repetitions~\cite{Vaart1998}. In what follows, we drop the dependences on the prior and the parameter true-value in the expressions for the MSE, Eqs.~(\ref{Eq:Av_MSE}) and (\ref{Eq:MSE}), respectively. However, let us emphasise that these manifest the fundamental difference between Bayesian and frequentist approaches to inference~\cite{Kay1993}.

	\subsection{Bounds on performance}
	Importantly, both the \emph{average MSE}~\eref{Eq:Av_MSE} and the \emph{local MSE} \eref{Eq:MSE} can be fundamentally lower-bounded within the Bayesian and frequentist settings, respectively. In what follows, we describe the corresponding bounds for an abstract $\theta$-parametrised probability distribution $p_\theta(x)$ of a random variable $X$, which in the quantum setting describes the outcome of a general measurement $\{M_x\}$ (a set with all $M_x\ge0$ and $\sum_x\,M_x=\openone$), i.e.~labels the element of a positive-operator-valued measure. Although any measurement performed continuously in time provides also an example of $p_\theta(x)\equiv P(D_t|\theta)$ with the $X$-variable representing then the full data $D_t$, we avoid such a notation, as we will use both the Bayesian and frequentist bounds to benchmark the performance of a \emph{strong single-shot} measurement, assessed as an alternative to the \emph{non-demolition time-continuous} detection of emitted photons.

	\subsubsection{Cram\'er-Rao bounds on local performance}
	\label{sec:CRBs}
	For any \emph{(locally) unbiased} estimator\footnote{Any \emph{unbiased estimator}, for which the second term  in the second line of \eqnref{Eq:MSE} vanishes, constitutes trivially a  \emph{locally unbiased estimator} around $\theta$, as $\av{\est{\theta}}{p_\theta}=\theta\;\implies\;\partial_\theta\,\av{\est{\theta}}{p_\theta}=1$.}, i.e.~satisfying $\partial_\theta\,\av{\est{\theta}}{p_\theta}=1$ at a specific true value of the parameter $\theta$, the local MSE \eref{Eq:MSE} is generally lower-limited by the \emph{Cram\'er-Rao bound}~\cite{Kay1993}:
	\begin{equation} \label{Eq:CRB}
		\Delta^2 \est{\theta} \;\underset{\text{\tiny{unb.}}}{\geq}\; \frac{1}{\nu}\,\frac{1}{F[p_\theta]} \, ,
	\end{equation}
	where $F[p_\theta]\eqdef\sum_x p_\theta(x) [\partial_\theta\ln p_\theta(x)]^2$ is the \emph{Fisher information} (FI) of the probability density function $p_\theta$. The local bound \eref{Eq:CRB} is guaranteed to be saturated in the asymptotic limit of independent repetitions, $\nu\to\infty$, also (under certain regularity conditions) by the optimal Bayesian estimator $\estB{\theta}$ with the mean and variance of the posterior distribution converging then to the true $\theta$ and $(\nu F[p_\theta])^{-1}$, respectively~\cite{Vaart1998}.   

	In the quantum regime, in order to obtain a fundamental bound on the local MSE that is determined solely by the quantum state $\rho_\theta$ in $p_\theta = \trace(\rho_\theta M_x)$, one should minimise \eqnref{Eq:CRB} over all measurement strategies, i.e.~maximise the FI over all quantum measurements $\{M_x\}$. In this way, one obtains the \emph{quantum Fisher information} (QFI), $F_{\rm Q} [\rho_\theta]\eqdef\max_{\{M_x\}}F[p_\theta]=\Tr{\rho_\theta\,L^2}$ with the (symmetric-logarithmic derivative) operator $L$ constituting the solution to the equation $\partial_\theta\rho_\theta=\frac{1}{2}(L\rho_\theta+\rho_\theta L)$~\cite{Braunstein1994}. This then leads to the \emph{quantum Cram\'er-Rao bound}:
	\begin{equation} \label{Eq:QCRB}
		\Delta^2 \est{\theta} \;\underset{\text{\tiny{unb.}}}{\geq}\; \frac{1}{\nu}\,\frac{1}{F_{\rm Q} [\rho_\theta]}\, ,
	\end{equation}
	which is again guaranteed to be attainable by an (asymptotically) unbiased estimator in the limit of many independent repetitions, $\nu\to\infty$. Still, it applies to any (locally) unbiased estimator for any $\nu$ (in particular, also in the single-shot scenario with $\nu=1$), constituting the ultimate bound dictated by the quantum mechanics.

	\subsubsection{Van Trees bounds on average performance}
	\label{sec:BCRBs}
	Within the Bayesian setting, it is the average MSE \eref{Eq:Av_MSE} that is the correct figure of merit, being defined with respect to a particular prior distribution, e.g.~the one of \eqnref{Eq:Prior}. As a result, any appropriate bound on estimation precision must include the information about the parameter present in the prior distribution, while not prioritising any particular parameter value.
	Here, we employ the \emph{van Trees bound} that reads~\cite{van2007bayesian}:
	\begin{align} \label{Eq:Van_trees}
		\av{\Delta^2 \est{\theta}} & \geq \frac{1}{F[P(\theta)] + \int\! \dd \theta\, P(\theta)\, F[p_{\theta}]},
	\end{align}
	where the first term in the denominator is the FI of the prior distribution, i.e.~$F[P(\theta)]=\int\dd\theta P(\theta)[\partial_\theta\ln P(\theta)]^2$, while the second term is the FI of the outcome distribution for a given value of $\theta$ that, in stark contrast to \eqnref{Eq:CRB}, must also be averaged over the prior.

	Similarly to the local case, in order to obtain the quantum generalisation of \eqnref{Eq:Van_trees}, one should minimise it over all measurement strategies, which results in replacing the FI by the QFI defined as before, and the \emph{quantum van Trees bound} reads
	\begin{align} \label{Eq:Q_Van_trees}
		\av{\Delta^2 \est{\theta}} & \geq \frac{1}{F[P(\theta)] + \int\! \dd \theta \, P(\theta) F_{\rm Q} [\rho_{\theta}]} \, .
	\end{align}
	Importantly, both van Trees bounds \erefs{Eq:Van_trees}{Eq:Q_Van_trees} apply to more realistic scenarios of quantum sensing, in which not only the unbiasedness of the estimator (or the asymptotic statistics, $\nu\to\infty$) is not required, but also the estimation is no longer performed around an \emph{a priori} known value of the parameter, as the prior distribution specifies now the effective range of $\theta$ to be considered.
	
	\subsubsection{Computing the quantum Fisher Information}
	\label{sec:qfi}
	We will be interested in computing the QFI of the density matrix representing the real-time state of our system obtained by solving either the ensemble \eref{Eq:master_full} or conditional \eref{Eq:Detector_stoch} master equation describing the dynamics. As before, this density matrix is taken to be supported by a truncated Fock space for both the cavity and mechanical modes, large enough to incorporate the complete evolution to high accuracy. However, no matter the origin of the quantum state of interest, its QFI can be conveniently evaluated with help  of the Bures distance, which is defined as~\cite{Yuan2017}:
	\begin{align} \label{Eq:Bures}
		d^2_{\rm B} (\rho_1 , \rho_2) & \eqdef  2 \left(1 - \sqrt{f(\rho_1 , \rho_2)} \right) \, ,
	\end{align}
	where $f(\rho_1 , \rho_2)$ is the quantum fidelity
		\begin{align} \label{Eq:Fidelity}
			f(\rho_1 , \rho_2) & \eqdef \Tr{\sqrt{ \sqrt{\rho_1} \rho_2 \sqrt{\rho_1}}}^2 \, .
		\end{align}
	Then, the QFI with respect to a parameter $\theta$ is related to the Bures distance by \cite{Zhou2019}
	\begin{align} \label{Eq:QFI}
		F_Q [ \rho_{\theta}] & = \lim_{\delta\theta\to0}\frac{d^2_{\rm B} \!\left(\rho_{\theta-\frac{\delta\theta}{2}} , \rho_{\theta + \frac{\delta\theta}{2}}\right)}{\delta\theta^2} \, .
	\end{align}
	In our case, we evaluate the right-hand side of \eqnref{Eq:QFI} by numerically integrating the master equation \eref{Eq:master_full} or \eref{Eq:Detector_stoch} (given a particular photon-click pattern in the latter case) for the parameter value $\theta$, as well as its small perturbation $\theta + \delta \theta$, where $\delta \theta$ we choose small enough to ensure an accurate determination of the QFI in \eqnref{Eq:QFI}, but large enough to provide numerical stability.
	
	\begin{figure*}[t] 
		\centering
		\includegraphics[width=0.99\linewidth]{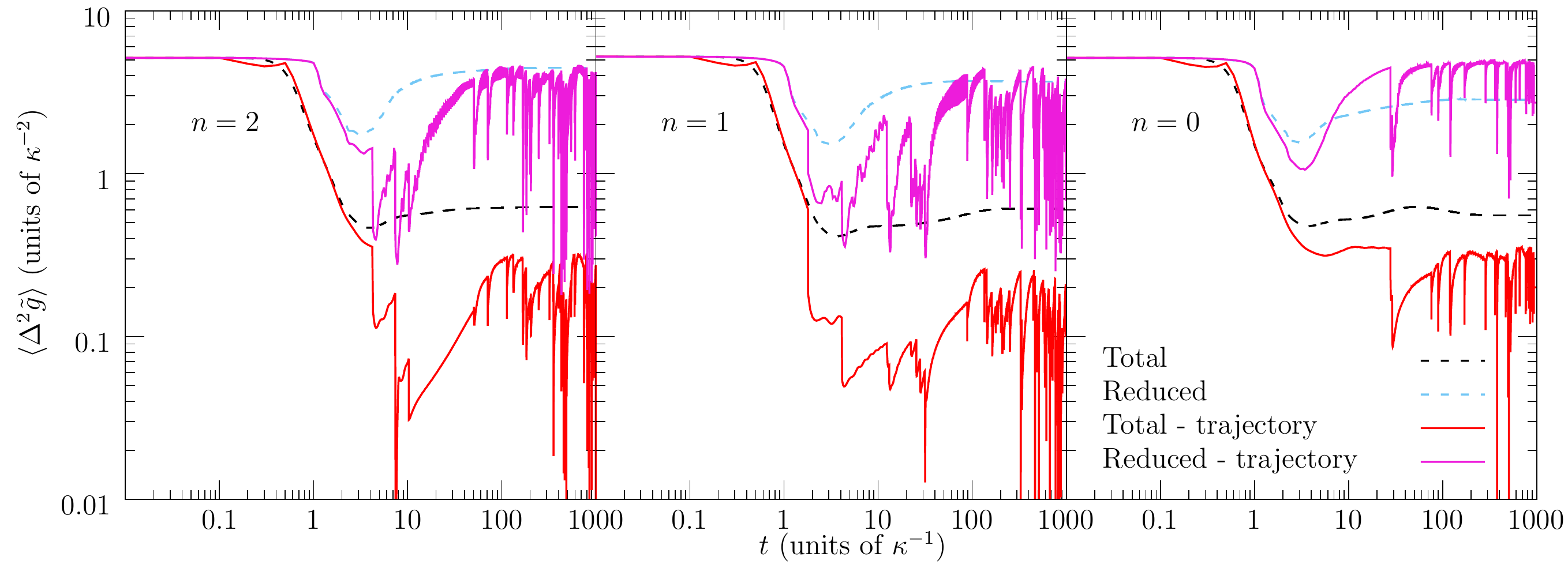}
		\caption{\textbf{Best average sensing performance of a single-shot measurement.}
			The minimal average estimation error as indicated by the Van Trees bound is shown for both the ensemble and along a specific trajectory generated for each of the detuning regimes.  The photon clicks cause significant drops in the bound, although they quickly increase back to a value following a similar trend to the ensemble.  We also show the bound of a reduced state with the mechanics traced out, for the same trajectory as the corresponding unreduced state.  By doing so, the bound is significantly increased, indicating a large part of the information is only accessible from the mechanics.  In all cases, the bound approaches a steady state, meaning no further information can be obtained by waiting longer.
		}
		\label{Fig:van_Trees}
	\end{figure*}

	\subsection{Ultimate average performance of a single-shot measurement}
	Similarly to the analysis of entanglement in \secref{sec:Open_dynamics} and \figref{Fig:Entanglement_open}, we would like to investigate the impact of conditioning -- monitoring the photon-click trajectory -- on the sensing capabilities of our optomechanical system. For that reason, we resort to the quantum van Trees bound \eref{Eq:Q_Van_trees} and compute it (see \figref{Fig:van_Trees}) for the system evolving according to the ensemble (\emph{dashed}) and conditional (\emph{solid}) dynamics, i.e.~Eqs.~\eref{Eq:master_full} and \eref{Eq:Detector_stoch} respectively, of which the latter is presented for a particular representative photon-click trajectory. 

	In order to compute the QFI in each case, we use the method described in the previous subsection, where as the estimated parameter we choose the coupling strength, i.e.~$\theta\equiv g$. However, similar analysis can be performed for other parameters affecting the photon-click pattern, e.g.~for the mechanical frequency $\omegaM$, as shown in \appref{app:sensing_simple}.

	Crucially, the van Trees bounds in \figref{Fig:van_Trees} represent the ultimate limit on the average MSE \eref{Eq:Av_MSE} for the most general single-shot measurement performed at a given time $t$. In particular, in case of the conditional evolution they ignore the information about the parameter within the click-pattern itself, but rather indicate the instantaneous sensing capabilities of the system.  However, the required strong measurement would have to be performed in principle on both the cavity and mechanical modes. That is why, we also present the bounds obtained after tracing out the mechanics, as even in an idealised setting only the optical mode should be considered accessible.
	
	We observe that, as in the case of negativity in \figref{Fig:Entanglement_open}, the conditional dynamics exhibits enhanced sensing capabilities at quantum jumps compared to the ensemble evolution.  Although both corresponding bounds on the average MSE reach its minimum early in the evolution, before increasing afterwards and finally reaching the steady state of the system, the trajectory-based bound typically drops significantly after each photon-detection event. Still, it tends towards a steady state in between the photon emissions, but the exact value of the average MSE being approached is different to the one emerging for ensemble dynamics. Moreover, the non-linearity of the interaction enhances the sensing capabilities when the system is continuously monitored. Although for the cascade ($n=2$) and blockade ($n=1$) regimes of detuning the gain may seem insignificant for the ensemble dynamics when comparing these with the case of on-resonance driving ($n=0$), in case of conditional dynamics (also in the reduced case of the cavity mode only) much smaller values of the average MSE may be reachable when $n=1$ or $n=2$.

	Importantly, any of the bounds depicted in \figref{Fig:van_Trees} precludes the average MSE to be vanishing as $t\to\infty$, as the existence of a stationary state limits the performance of the optimal single-shot measurement at long times, irrespectively whether ensemble or conditional dynamics is considered. As a consequence, when dealing with an optomechanical sensor operating in real time -- the main motivation of this work -- after the stationary state is reached, e.g.~$t\gtrsim100$ in \figref{Fig:van_Trees}, one should focus on extracting most efficiently the value of the parameter solely from the photon-click pattern. This is because the photon-click trajectory keeps containing more and more information as time progresses and more photons are being emitted, which must inevitably outrun the information extractable from the system that is fundamentally limited, as shown in \figref{Fig:van_Trees}. Hence, although one may try to adapt the van Trees bounds \erefs{Eq:Van_trees}{Eq:Q_Van_trees} to incorporate the continuous quantum measurement (here, photon-counting) -- see e.g.~\cite{albarelli_ultimate_2017} for the adaptation of the local bounds \erefs{Eq:CRB}{Eq:QCRB} (for homodyne detection) -- from the perspective of the real-time sensing scenario considered here is unnecessary. The optimal sensing performance is always attained along a single experimental run by engineering the system, so that the collected data of photon emissions, $D_t$, yields an average \eref{Eq:Av_MSE} or local \eref{Eq:MSE} MSE that drops most rapidly with time $t$. In what follows, we demonstrate that non-linear effects and, in particular, the correct choice of detuning, play a crucial role in this respect.

	\section{Sensing from photon-clicks}
	\label{sec:sensing}

	As motivated by the previous section, our goal is to most accurately infer the parameter being sensed, here the coupling constant $g$ (see also \appref{app:sensing_simple} for inference of the mechanical frequency $\omegaM$), from the photon-click pattern being observed. We would like to investigate how the non-classical statistics of photon emissions affect the sensing performance, in particular, compare the previously defined three distinct cases of detuning:~on resonance driving ($n=0$) in which the emitted photons do not exhibit any correlations, blockade regime ($n=1$) in which photons bunch at moderate times (the blue region in \figref{Fig:Samp}), and the cascade regime ($n=2$) in which the photons prefer to be emitted in pairs and, hence, bunch at very short timescales (the red region in \figref{Fig:Samp}). 

	However, as each of these scenarios requires the detuning parameter to be set to $\Delta = - n g^2 / \omegaM$ for an adequate $n$, see \eqnref{Eq:energy_levels}, these can be unambiguously compared only for a particular value of $g$. That is why, rather than considering the complete Bayesian setting and drawing the true value of $g$ from a prior distribution, what would require different sets of values for $\Delta$ to be chosen depending on each true value of $g$, we consider local estimation of $g$ around a given \emph{fixed} value. 

	Nonetheless, we importantly disallow the true parameter value to be \emph{a priori} available when constructing the real-time estimator, in particular, we use the optimal Bayesian estimator described in \secref{subsec:bayinfer} that still assumes the prior distribution \eref{Eq:Prior} to be valid. As an example, we choose in our simulations the true value of $g=4$ (in units of $\kappa$), while the estimator expects it to be distributed in the range $g\in[2,10]$ according to \eqnref{Eq:Prior} with $\alpha=-1000$.

	\begin{figure*}[ht!] 
		\centering
		\includegraphics[width=0.98\linewidth]{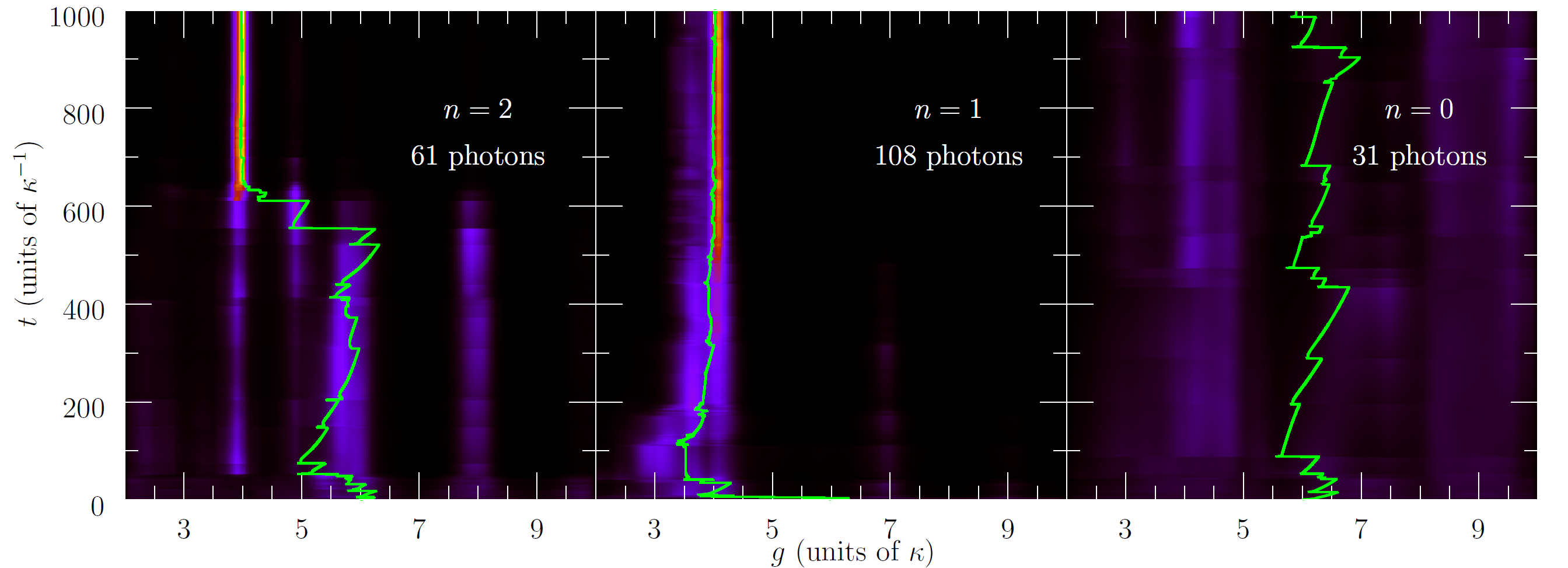}
		\includegraphics[width=0.98\linewidth]{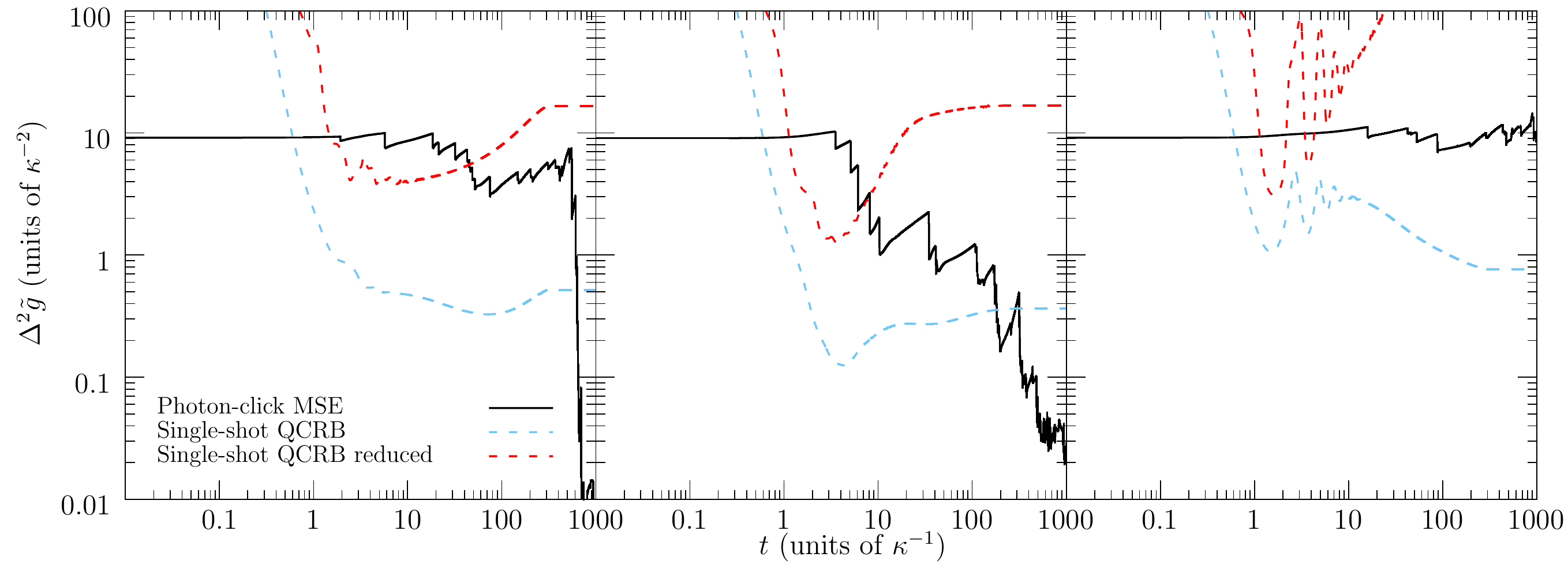}
		\caption{ 
			\textbf{Sensing performance from a photon-click pattern}.
			In the top row, we plot the posterior distribution over the optomechanical coupling, $P(g|D_t)$ along with its mean marked in green, for typical photon-click trajectories $D_t$ generated over a large period of time, beginning with a flat prior of the form \eref{Eq:Prior} (with $\alpha = -1000$) that assumes $g\in[2,10]$.  Underneath, we show the corresponding MSE of the optimal Bayesian estimator $\estB{g}(D_t)$, i.e.~the mean of the posterior, along with the local bounds of single-shot measurement for both the total and reduced state of the system. The trajectories are generated with a true value of $g=4$ (in units of $\kappa$). For detuning with $n=2$ (cascade regime), the estimator initially struggles to infer the parameter, but once a sufficient number of photons are detected its MSE decreases extremely rapidly.  For the $n=1$ case (blockade regime), we observe a steady decrease in the estimator's uncertainty that begins earlier, due to the higher number of photons being then emitted.  For on-resonance driving ($n=0$), sensing turns out to be impossible as, not only does a typical photon-click trajectory contain a modest number of photon emissions, but also these contain then hardly any information about $g$.  All the system characteristics are set as in \figref{Fig:Entanglement_open}, while $g$ is varied only in the construction of the estimator.
		}
		\label{Fig:Bayes_regimes}
	\end{figure*}

	\subsection{Single-shot measurement as a benchmark}
	\label{sec:single-shot_meas}	
	Hence, as we are considering here the sensing performance for a specific value of $g$, we should judge our results by rather using the  Cram\'er-Rao bounds \erefs{Eq:CRB}{Eq:QCRB} on the local MSE \eref{Eq:MSE}. Importantly, these apply also to the single-shot scenario ($\nu=1$) and can still be compared with the performance of the optimal Bayesian estimator, $\estB{\theta}$, as long as the latter can be considered unbiased. 
	That is why, in \figref{Fig:Bayes_regimes} (dashed lines in the bottom plots) we include the quantum Cram\'er-Rao bounds \eref{Eq:QCRB} evaluated for the ensemble-averaged state of the system. Similarly to the bounds presented in \figref{Fig:van_Trees}, these should be treated as a benchmark quantifying the instantaneous sensing capability of the system, if one allowed for the optimal strong measurement to be performed at a given time instance and ignored the information contained within the photon-click pattern, when inferring the (now, fixed) value of $g$. As before, we present these benchmarks also after tracing out the mechanical mode of the conditional state, as in reality it should be considered inaccessible.
	
	\subsection{Sensing the true value of $g$}
	Finally, we demonstrate the superiority of sensing capabilities for the optomechanical system when inferring the parameter from photon-clicks. Moreover, being interested in exploiting the non-linear effects, we consider again the three distinct regimes of cascade ($n=2$), blockade ($n=1$) and on-resonance ($n=0$) detuning, which are compared against one another horizontally in \figref{Fig:Bayes_regimes}. In particular, within the top plots of \figref{Fig:Bayes_regimes} we present explicitly the evolution with time of the posterior distribution $P(g|D_t)$ along a representative photon-click trajectory, together with its corresponding mean, i.e.~the optimal Bayesian estimator $\estB{g}(D_t)$. Within the bottom plots of \figref{Fig:Bayes_regimes}, we then plot the corresponding MSE \eref{Eq:MSE} achieved by $\estB{g}(D_t)$ in time for each of the three non-linear regimes, while comparing it against the single-shot measurement benchmarks described above. All the properties of the optomechanical system are set the same as in \secref{sec:Open_dynamics}, which crucially yield the true value of $g=4$ (in units of $\kappa$).  

	Although for the cascade regime ($n=2$) a typical photon-click trajectory does not contain that many photons (here, 61), once a sufficient number of photons are emitted the MSE rapidly drops thanks to their strong non-classical correlations. As a result, this turns out to be the best scenario within the three regimes considered here, given the particular trajectories generated for each of the three cases:~it yields then the smallest error at the final time, $t=1000$ (in units of $1/\kappa$), of the simulation.
	However, as is clear from the plot of the posterior distribution in \figref{Fig:Bayes_regimes}, its corresponding estimator does not converge directly to the true parameter value -- in contrast to the $n=1$ case -- but rather requires significant time, given a relatively wide prior, for sufficient statistics of bunched photons to be detected, so that then (here, at $t\approx600$) the correct $g$-value is resolved and the MSE rapidly drops. Unless the prior is chosen to be narrower (e.g.~$g\in[2,5]$ to avoid the incorrect values emergent in \figref{Fig:Bayes_regimes}), such a waiting time may be large and is not predictable. As we show explicitly in App.~\ref{app:Robust}, where we present the MSEs of \figref{Fig:Bayes_regimes} for $n=1,2$ after being averaged further over typical photon-click trajectories arising in each regime, such averaging results in $\Delta^2\tilde{g}$ to diminish quicker in the blockade regime with $n=1$, while the "sudden-drop" feature is nevertheless exhibited along the trajectories observed in the cascade regime with $n=2$.

	On the other hand, the blockade regime ($n=1$) yields good overall performance, as despite the emitted photons being less correlated, their abundance is higher (here, 108 emissions for a typical photon-click trajectory). Still, let us emphasise that in both cases the sensitivity constantly grows with time, with more and more photons being detected, clearly surpassing the single-shot measurement benchmarks that are limited by the steady-state behaviour. Note that these may be unambiguously compared at long times at which $\estB{g}(D_t)$ becomes unbiased, as may be directly verified by inspecting the posterior distributions for $n=1,2$, which from $t\gtrsim600$ narrowly peak around the true value of $g$~\cite{Vaart1998}.

	It may have been expected that on-resonance driving ($n=0$) turns out to be useless for the sensing task considered, as in this regime not only are the photons rarely emitted (here, 31 emissions in an exemplary photon-click pattern), but also they hardly interact with the mechanical oscillator~\cite{Aspelmeyer2014}, carrying little information about $g$. However, one should bear in mind that this is a consequence of the particular characteristics of the sensor being chosen to focus on its non-linear properties. In particular, in \appref{app:sensing_simple}, we also demonstrate that for stronger driving but weaker coupling (closer to the linear regime) also the on-resonance regime yields unconstrained sensing capabilities with time, when considering either $\omegaM$ or $g$ as the parameter to be inferred.
	
	Let us also comment that one may interpret the superiority of the cascade ($n=2$) over the blockade ($n=1$) regimes by the fact that the former yields photon-click patterns that reveal more information about the parameter per photon~\cite{Clark2019}. As such, at long times once a sufficient number of photons is emitted on average also for the $n=2$ case, they are always capable of beating the sensitivity achieved by the $n=1$ trajectory, despite higher abundance of photon-clicks within the latter. In a similar way, if one was to consider ``higher-order'' cascade regimes with $n>2$ in \eqnref{Eq:energy_levels} (photons preferring to be emitted in triplets, quadruplet etc.), one should expect further enhancement thanks to even stronger inter-photon correlations. However, the time it would take for such an enhancement to matter could be large, as the rate at which such multi-photon groups are emitted would dramatically fall.
	
	Finally, let us emphasise that in order to investigate the impact of non-classical photon correlations, we have assumed here idealistically that the sensor is \emph{a priori} calibrated (e.g.~by a third party) with its detuning $\Delta$ being set to satisfy perfectly either the blockade or the two-photon cascade conditions. In practice, however, such calibration would require the value of the sensed parameter (here, $g$) to also be known in advance, what suggests that it would need to be performed in an adaptive manner along the sensing procedure. Nonetheless, this does not constitute a fundamental problem, as the sensor exhibits robustness to deviations of $\Delta$ from a desired value, i.e.~it still then yields errors that eventually diminish with time, as shown explicitly in \appref{app:Robust}.

	\section{Conclusions}
	\label{sec:conclusions}
	We have shown that recent experimental breakthroughs demonstrating the ability to measure and control optomechanical sensors in real time, as well as reaching regimes of non-linear interactions between optical and mechanical modes, open doors for a new generation of optomechanical sensors that operate continuously in time. 

	In particular, by considering the canonical optomechanical setup we have demonstrated with help of numerical methods, which importantly allowed us to incorporate relevant decoherence mechanism into the non-linear sensor dynamics, that by continuously driving the system into the stationary state in balance with the dissipation processes and detecting the photons being emitted, one can gain precise information about external parameters perturbing the sensor. 

	Moreover, by correctly adjusting the (red) detuning of the driving field the non-linear effects can be enhanced, so that thanks to non-classical correlations (bunching) between the photon-clicks the sensor performance is improved. Although such an approach requires fine-tuning of optical driving parameters that may vary with the parameter being sensed, we believe that our work paves the way for future proposals of schemes that involve active feedback~\cite{sudhir_appearance_2017,Setter2018,mason_continuous_2019,magrini_real-time_2021}, so that optimal non-linear characteristics can be maintained during the sensing process, while altering the driving field `on-the-fly'.
	
	Although we have presented results for sensing the optomechanical coupling in \secref{sec:sensing}, our methods are easily extendible to other parameters.  Moreover, at the price of exploiting less the non-classical characteristics of detected photons, one may also consider regimes of stronger driving that lead to higher rates of photon-counts and, hence, more data. Both of these factors are considered in App.~\ref{app:sensing_simple}, where we infer the mechanical frequency of the oscillator by our methods, with the Bayesian inference protocol coping efficiently with higher rates of detection. As such, we believe that our sensing scheme can be implemented with a broad spectrum of optomechanical-like devices despite their different characteristics, as well as other cavity-\cite{Reiserer2015,peyronel_quantum_2012,Mohl2020} or waveguide-based~\cite{prasad_correlating_2020} systems with tunable non-linearity of light-matter interactions~\cite{chang_quantum_2014}.
	
	A limitation of our results is the speed at which data can be analysed. As discussed, the bottleneck in generating the posterior distribution comes from acquiring the likelihood function. Approaches previously considered, such as using the Metropolis-Hastings algorithm~\cite{Gammelmark2013}, are not expected to be particularly useful in our sensing scenario, as they are generally tailored instead to problems with in large, high-dimensional parameter spaces. Nonetheless, exploring other pathways to improve the speed of Bayesian-inference methods we propose is therefore expected to be an interesting development of our work.
	
	From the theoretical perspective, let us note that the ultimate bounds on average precision for a single-shot destructive measurement, which we have used within our work as a benchmark, could be in principle generalised to include the continuous inflow of information gained from photon-counting~\cite{Gammelmark2014,albarelli_ultimate_2017}, or even further limited based solely on the properties of the decoherence~\cite{amoros-binefa_noisy_2021}. Such an analysis 
	would allow us to verify the optimality of photon-detection as a quantum measurement, which we leave open for the future.

	\section*{Acknowledgements}
	We thank Witlef Wieczorek and Micha\l{} Parniak for many useful comments. This research was supported by the Foundation for Polish Science within the “Quantum Optical Technologies” project carried out within the International Research Agendas programme cofinanced by the European Union under the European Regional Development Fund, and also financed by the project C’MON-QSENS! that is supported by the National Science Centre (2019/32/Z/ST2/00026), Poland under QuantERA, which has received funding from the European Union's Horizon 2020 research and innovation programme under grant agreement no 731473.

	\appendix
	%
	\addtocontents{toc}{\setlength{\cftsecnumwidth}{16ex}}
	\addtocontents{toc}{\setlength{\cftsubsecnumwidth}{16ex}}

	\section{Closed and fully conditional state evolution}
	\label{ap:app1}
	Here we show how a quantum state evolves according to \eqnref{Eq:Non_unitary_evolution_operator} and as such find the purity of the resulting reduced state used in \figref{Fig:Entanglement_closed}, following similar methods as in Ref.~\cite{qvarfort_gravimetry_2018}.
	
	\subsection{Dynamics}
	We assume we begin in an initial coherent state for both cavity and mechanics of
	\begin{align} \label{Eq:Initial_state}
		\ket{\psi (0)} = & \cav{\ket{\alpha}} \mech{\ket{\beta}} \, .
	\end{align}
	Applying \eqnref{Eq:Non_unitary_evolution_operator} to \eqnref{Eq:Initial_state} we find
	\begin{align}
		\ket{\psi (t)} = & {\rm e}^{-\frac{|\tilde{\alpha}|^2}{2}} \sum\limits_{n=0}^\infty \frac{\tilde{\alpha}^n}{\sqrt{n!}} {\rm e}^{{\rm i} k^2 n^2 \left(t' - \sin(t')\right)} \nonumber \\
		& \times {\rm e}^{k n \left(\beta \eta - \beta^* \eta^*\right)/2} \cav{\ket{n}} \mech{\ket{\phi_n}} \, ,
	\end{align}
	where $\ket{\phi_n} = \ket{\beta {\rm e}^{-{\rm i} t'} + k n \eta}$ is a coherent state and $\tilde{\alpha} = \alpha {\rm e}^{-{\rm i} \tilde{r} t'}$.  The corresponding density matrix $\rho (t) = \ket{\psi (t)} \bra{\psi (t)}$ is hence
	\begin{align}
		\rho = & {\rm e}^{-| \tilde{\alpha} |^2} \sum\limits_{m,n = 0}^\infty \frac{\tilde{\alpha}^n (\tilde{\alpha}^*)^m}{\sqrt{m!} \sqrt{n!}} {\rm e}^{{\rm i} k^2 \left(n^2 - m^2\right) \left(t' - \sin(t')\right)} \nonumber \\
		& \times {\rm e}^{k (n-m) \left(\beta \eta - \beta^* \eta^*\right)/2} \ket{n} \ket{\phi_n} \bra{\phi_m}   \bra{m} \, ,
	\end{align}
	where we have dropped the subscript labels on the states as it is clear which belongs to which space.  We now wish to consider the reduced system with the mechanics traced out.  The density matrix of the reduced state $\cav{\rho} = \mech{\rm Tr} (\rho)$ is then
	\begin{align} \label{Eq:reduced_density_matrix}
		\cav{\rho}(t) = & {\rm e}^{-|\tilde{\alpha}|^2} \sum\limits_{m,n = 0}^\infty \frac{\tilde{\alpha}^n (\tilde{\alpha}^*)^m}{\sqrt{m!} \sqrt{n!}} \nonumber \\
		& \times  {\rm e}^{{\rm i} k^2 \left(n^2 - m^2\right) \left(t' - \sin(t')\right)} \nonumber \\
		& \times {\rm e}^{k (n-m) \left(\beta \eta - \beta^* \eta^*\right)/2} \nonumber \\
		& \times {\rm e}^{-\frac{|\phi_n|^2}{2} - \frac{|\phi_m |^2}{2} + \phi_n \phi_m^*} \ket{n}{\bra{m}} \, .
	\end{align}
	We can now use this density matrix to find the properties of the reduced state.
	
	\subsection{Purity}
	The purity of a density matrix is defined to be $P = {\rm Tr}({\rho^2})$.  Using the density matrix in \eqnref{Eq:reduced_density_matrix}, we find
	\begin{align}
		\cav{\rho}^2 (t) = & {\rm e}^{-2 |\tilde{\alpha}|^2} \sum\limits_{m,n,n'=0}^\infty \frac{\tilde{\alpha}^n (\tilde{\alpha}^*)^m |\tilde{\alpha}|^{2 n'}}{\sqrt{n! m!} n'!} \nonumber \\
		& \times {\rm e}^{{\rm i} k^2 \left(n^2 - m^2 \right) \left(t' - \sin(t')\right)} \nonumber \\
		& \times {\rm e}^{k (n-m) \left(\beta \eta - \beta^* \eta^*\right)/2} \nonumber \\
		& \times {\rm e}^{-\frac{|\phi_n|^2}{2} - \frac{|\phi_m|^2}{2} - |\phi_{n'}|^2 + \phi_n \phi_{n'}^* + \phi_{n'}\phi_m^*} \nonumber \\
		& \times \ket{n} {\bra{m}} \, . 
	\end{align}
	The purity of the system is hence
	\begin{align}
		P = & {\rm e}^{- 2 |\tilde{\alpha}|^2} \sum\limits_{n,n' = 0}^\infty \frac{|\tilde{\alpha}|^{2 (n + n')}}{n! n'!} \nonumber \\
		& \times {\rm e}^{-\left| \phi_n \right|^2 - \left| \phi_{n'} \right|^2 + 2 {\rm Re}\left(\phi_n \phi_{n'}^*\right)} \, .
	\end{align}
	The size of $|\alpha|^2$ will determine how high $n,n'$ need to be summed to in order to obtain good results, but this value can be obtained for large $n,n'$ easily numerically.

	\section{On-resonance sensing from photon-clicks closer to the linear regime}
	\label{app:sensing_simple}
	
		\begin{figure*}[t] 
		\centering
		\includegraphics[width=0.99\linewidth]{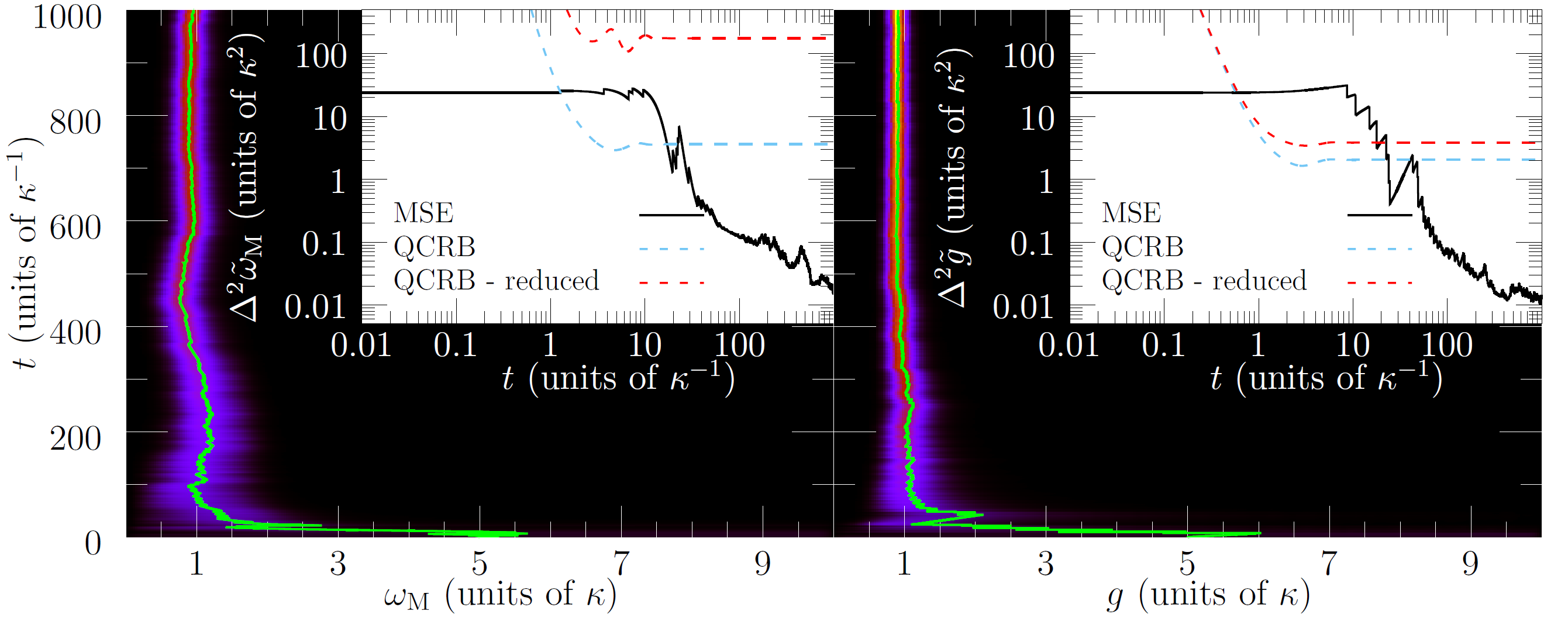}
		\caption{
			{\textbf{Bayesian inference of $g$ and $\omegaM$ closer to linear regime.}}  We take a prior distribution of the form of \eqnref{Eq:Prior} with $\theta_{\rm min} = 0$ and $\theta_{\rm max} = 10$ and with $\alpha = -1000$.  In both cases, we see that the posterior distribution focuses on the true value increasingly sharp as a function of time.  To see this, we also plot the MSE $\Delta^2 g$ and $\Delta^2 \omegaM$.  In both cases we generate a trajectory from a system with parameters $\Delta = 0$, $\Omega = g = \omegaM = \gamma = \kappa$, $\kappad = 0.9 \kappa$, $\kappal = 0.1 \kappa$ and $\bar{m} = 1$.  Here, there are $\sim 300$ photons emitted, which is much higher than the results within \secref{sec:sensing}, yet the MSE at large times is still comparable.
		}
		\label{Fig:Bayes_1}
	\end{figure*}
	We now consider a full Bayesian analysis of the photon statistics of a quantum trajectory to infer the value of $g$ or $\omegaM$, using the methods described in \secref{sec:bayesian}, for a system prepared closer to the linear regime of optomechanics.  In particular, we now set $\Delta = 0$, $\Omega = g = \omegaM = \gamma = \kappa$, with again $\kappad = 0.9 \kappa$, $\kappal = 0.1 \kappa$ and $\bar{m} = 1$.  While we still do not perform linearisation of the optomechanical Hamiltonian and its Langevin equations, these parameters are out of the sideband-resolved regime, and much closer to the linear regime---they do not satisfy the inequalities in Eq.~(\ref{Eq:Non-lin}) -- due to the stronger driving and weaker optomechanical coupling. Nevertheless, there exist quantum signatures of the dynamics that can 
	still be observed within such a regime~\cite{Ludwig2008}.
	
	We show results for inference with these parameters in \figref{Fig:Bayes_1}.  We see that the Bayesian inference is capable of predicting the correct value of $g$ or $\omegaM$ to a high level of accuracy, thus demonstrating that our results are not dependent on being in a special regime of optomechanics.  Indeed we find the photon statistics can provide information about unknown parameters if either there is enough of them or they are correlated.
	
	For comparison, we also compare to the local bound set by Cram\'er-Rao bound, as we did in \secref{sec:sensing}.  As for our main results, this bound initially decreases rapidly, before settling at a stationary value.  We see that for measuring $g$, within this regime there is much less difference between the bound for the total and reduced state.  However, when considering $\omegaM$, this difference is much more substantial.  Indeed, almost all the information about $\omegaM$ is contained within the mechanics.  Nevertheless, it is then surprising to see that the inference from the photon-clicks is capable of measuring $\omegaM$ nearly as well as for $g$.  In both cases, we again see that with a sufficient waiting time, the Bayesian inference from continuous photon-counting can beat a single-shot measurement.

	We note that the parameters chosen here, indicated in \figref{Fig:Bayes_1}, are not completely in-line with what is currently considered experimentally~\cite{Aspelmeyer2014}.  The parameters used here are chosen to minimise numerical complexity within a simple regime to obtain good results that demonstrate the capability of photon-counting for quantum sensing in optomechanics.
	
	\section{Robustness of sensing performance} 
	\label{app:Robust}
	In this appendix, we analyse in more detail theperformance of our sensing scheme, in order to draw attention to the robustness, as well as limitations, of our main results.

	\begin{figure*}[t] 
		\centering
		\includegraphics[width=0.99\linewidth]{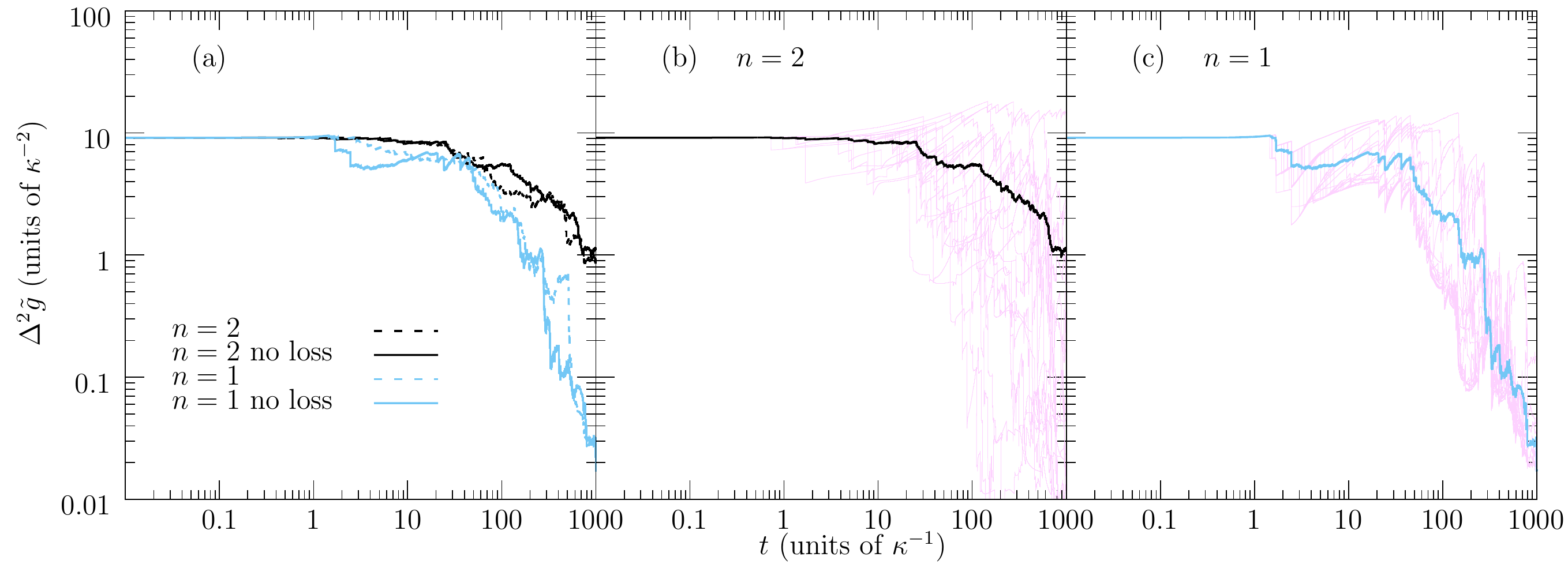}
		\caption{
			{\textbf{MSE averaged over photon-click trajectories for the $n=2$ and $n=1$ regimes.}}  (a) We obtain an average MSE by sampling 20 trajectories obtained within cascade ($n=2$) and blockade ($n=1$) regimes, considering in each case both the imperfect photon detection as in the main body (\emph{dashed lines}), as well as the perfect photon detection (\emph{solid lines}). In both cases, the trend is the same, thus demonstrating a robustness to small photon losses, while on average the behaviour is better for the $n=1$ regime.  The average MSEs for $n=2$ and $n=1$ are again plotted in (b) and (c), respectively, together with the MSEs obtained with each of the corresponding individual trajectories. The MSEs for $n=2$ clearly show larger fluctuations
			than those of $n=1$, while for each trajectory obtained for $n=2$, the MSE exhibits a steep drop (owing to two-photon correlations) at a sufficiently long time, as in \figref{Fig:Bayes_regimes} of the main text.
		}
		\label{Fig:Average_error}
	\end{figure*}

	\subsection{Averaging over different trajectories}
	Firstly, let us note that MSEs depicted in both Figs.~\ref{Fig:Bayes_regimes} and \ref{Fig:Bayes_1} show the results for single trajectories.  Hence, as each trajectory behaves differently, they may not necessarily be representative of the typical behaviour of the system.  In order to demonstrate that such behaviours are indeed representative of the dynamics, we present here in \figref{Fig:Average_error}(a) the two relevant ($n=2$ and $n=1$) MSEs of Fig.~\ref{Fig:Bayes_regimes}, however, after averaging them further over many, here 20, photon-click trajectories (the MSEs for each single trajectory used are also plotted in \figref{Fig:Average_error}(b) and (c), respectively). 

	Consistently, we observe that in both regimes of two-photon cascades, $n=2$, and photon blockade, $n=1$, the MSE decreases in time, but on average it is rather the $n=1$ regime that performs better -- given the same wide prior $g\in[2,10]$ assumed as in \figref{Fig:Bayes_regimes}. This is a result of a more consistent performance of trajectories in this regime, whereas the trajectories for $n=2$ lead to much larger variation of the MSE, as can be seen by comparing plots (b) and (c) in Fig.~\ref{Fig:Average_error}. Part of the reason for the larger performance spread in the $n=2$ case is the need for the data to possess sufficient bunching of the photon statistics, in order to certify that the parameter is within a range of values for which this strongly occurs.  As such, the `quality' of the trajectory becomes critical. This is manifested by the single-trajectory MSEs in Fig.~\ref{Fig:Average_error} (b) sharply dropping to zero thanks to the two-photon correlations after a sufficient time has passed, whose amount, however, is not well predictable -- yielding a worse MSE overall, once the averaging over trajectories is performed.
	
	We also compare in \figref{Fig:Average_error}(a) the trajectory-averaged MSEs obtained for perfect photon detection ($\kappad = \kappa$, $\kappal = 0$) with the case considered in the main body, in which a $10\%$ of photons is further lost ($\kappad = 0.9\kappa$, $\kappal=0.9\kappa$). This allows us to verify that, as there is little difference of performance between these two cases:~not only the scheme is robust to photon losses, but also the above conclusions remain valid also when $\kappal>0$.

	\subsection{The impact of imperfect detuning}
	Within the main text, in order to focus on the impact of non-classical correlations between the emitted photons on the sensor performance, we have assumed the detuning $\Delta$ to take the exact values that lead to particular regimes of photon statistics ($n=0,1,2$).  However, knowing the correct value of $\Delta$ to choose {\em a priori} is not feasible in practice, as it in principle requires also the  knowledge of the parameter being sensed. Here, we consider detunings chosen close to, but different from, the optimum values and find that, despite photon bunching/antibunching not being exactly satisfied, the evolution of the estimation error is only mildly affected.
	
\begin{figure}[t] 
		\centering
		\includegraphics[width=0.99\linewidth]{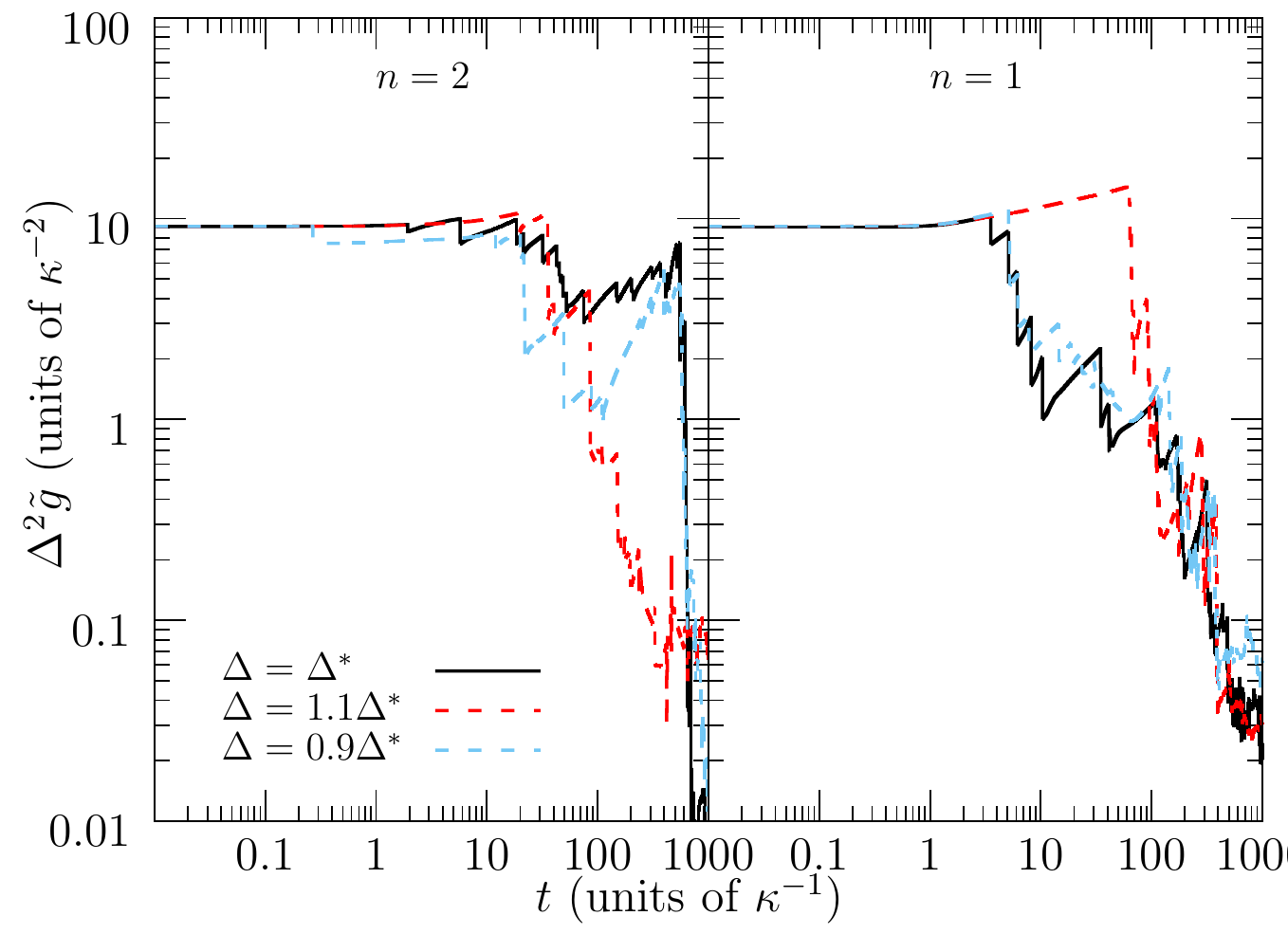}
		\caption{
			{\textbf{MSE for trajectories with perfect and imperfect detuning.} Exemplary trajectories for both the $n=2$ and $n=1$ regimes, compared to trajectories where the detuning is perturbed away from its ideal case.  In both cases, we find that the sensing is robust to such perturbations and still manages to sense the unknown parameter nearly as effectively.  All parameters are set as in Fig.~\ref{Fig:Bayes_regimes}, with only detunings being modified.}     
		}
		\label{Fig:Detuning_perturb}
	\end{figure}

	In Fig.~\ref{Fig:Detuning_perturb}, we compare the MSEs presented for $n=1$ and $n=2$ in Fig.~\ref{Fig:Bayes_regimes}, where the detuning is set to $\Delta^* = n g^2 / \omegaM$, to the ones obtained for imperfect detunings $\Delta = 0.9 \Delta^*$ and $\Delta = 1.1 \Delta^*$. In that latter two cases, new photon-click trajectories must be generated for them to also account also for the shift of $\Delta$, but nonetheless	the MSE behaves similarly to that of the ideal detuning. Hence, this demonstrates that the perfect calibration of detuning is not essential for the sensing performance.  The reason for this is that the photon statistics predicted by the detuning regimes are still present, only in a slightly weaker form. This can be seen by considering the energy landscape in \eqnref{Eq:energy_levels}.  While the energy separation in the $\ket{0} \rightarrow \ket{n}$ transition is no longer zero for imperfect detuning, it is still reduced and thus still favoured over other transitions, albeit in a slightly weaker sense.  Hence, even with the limitation of choosing the detuning incorrectly, our sensing scheme is capable of inferring the unknown parameter, while still benefiting from the -- now, imperfect -- non-classical correlations of the emitted photons.
	
	\bibliographystyle{myapsrev4-1}

	\bibliography{optomech_photoncount}
	
\end{document}